\documentclass[fleqn,usenatbib]{mnras}

\usepackage{newtxtext,newtxmath}
 
\usepackage{array}

\usepackage[T1]{fontenc}

\usepackage{graphicx}	
\usepackage{amsmath}	
\usepackage{amssymb}	

\title[Growth of early-type galaxies]{Late growth of early-type galaxies in low-z  massive clusters}

\author[Ribeiro et al.]{
A.L.B. Ribeiro$^{1}$, \thanks{E-mail: albr@uesc.br}
R.S. Nascimento$^{2}$,
D.F. Morell$^{1,3}$,
P.A.A. Lopes$^{4}$,
C.C. Dantas$^{5}$,
and M.H.S. Fonseca$^{4,1}$
\\
$^{1}$Laborat\'orio de Astrof\'isica Te\'orica e Observacional, Universidade Estadual de Santa Cruz, Ilh\'eus, BA 454650-000, Brazil\\
$^{2}$Laborat\'orio Nacional de Astrof\'isica/MCTI, Itajub\'a 37504-364, Brazil\\
$^{3}$Observat\'orio Nacional, Rio de Janeiro, RJ, Brazil\\
$^{4}$Observat\'orio do Valongo, Universidade Federal do Rio de Janeiro, RJ 20080-090, Brazil\\
$^{5}$ Instituto Nacional de Pesquisas Espaciais/MCTI, SP, Brazil\\
}

\date{Accepted XXX. Received YYY; in original form ZZZ}

\pubyear{2015}

\begin{document}
\label{firstpage}
\pagerange{\pageref{firstpage}--\pageref{lastpage}}
\maketitle

\begin{abstract}
We study a sample of 936 early-type galaxies (ETGs)  located in 48 low-z regular galaxy clusters with  $M_{200}\geq 10^{14}~ M_\odot$ at $z< 0.1$. 
We examine variations in the concentration index, radius, and  color gradient of ETGs as a function of their stellar mass and  loci in
the projected phase space (PPS) of the clusters.
We aim to understand the environmental influence on the growth of ETGs 
according to the time since infall into their host clusters.  Our analysis indicates a significant change in the behavior of the concentration index $C$ and color gradient 
around $M_{\ast} 
\approx 2\times 10^{11} ~M_\odot \equiv \tilde{M}_{\ast}$. 
Objects less massive than $ \tilde{M}_{\ast}$ present a slight growth of $C$ with $M_{\ast}$ with negative 
and approximately constant color gradients in all regions of the PPS. Objects more massive than $ \tilde{M}_{\ast}$
present a slight decrease of $C$ with $M_{\ast}$ with color gradients becoming less negative and approaching zero. 
We also find that objects more massive than $ \tilde{M}_{\ast}$, in all PPS regions, have smaller $R_{90}$ for a given $R_{50}$, suggesting a smaller external growth in these objects or even a shrinkage possibly due to tidal stripping.  
Finally, we estimate different dark matter fractions for galaxies in different regions of the PPS,
with the ancient satellites having the largest fractions, $f_{DM}\approx 65$\%.
These results favor a scenario where cluster ETGs  experience environmental influence the longer they remain and the deeper into the gravitational potential they lie, 
indicating a combination of tidal stripping + harassment, which predominate during infall, followed by mergers + feedback effects affecting 
the late growth of ancient satellites and BCGs.

\end{abstract}

\begin{keywords}
galaxies: evolution; galaxies: elliptical and lenticular, cD; galaxies: clusters
\end{keywords}



\section{Introduction}

Local early-type galaxies (ETGs) encompass objects that normally ceased their star formation,
have red colors, small amounts of cold gas
and dust, and which correspond morphologically to ellipticals  and
lenticulars \citep[e.g.][]{kauffmann2004environmental,blanton2009physical}. Some studies indicate that the formation of massive ETGs has occurred
in two phases \citep[e.g.][]{oser2010two}. At an early stage,
the gas collapses into dark matter halos and forms stars intensely for a
short time interval \citep[e.g.][]{thomas2005epochs,peng2010mass,conroy2015preventing}. The second phase involves mass accumulation through a
series of mergers \citep[e.g.][]{naab2009minor,feldmann2011hubble,johansson2012forming,huang2016characterizing} that enrich galaxies with
stars set \textit{ex-situ}. In the ${\rm \Lambda}$CDM cosmology,
structures form hierarchically, with  massive systems
forming through the addition of less massive objects. It means
that  massive galaxies may have accreted large fractions
of their current mass, with the accreted components deposited mainly in
its outermost parts.
On the contrary, low-mass galaxies are mostly made of
stars formed \textit{in-situ} \citep[e.g.][]{rodriguez2016stellar,pillepich2018first}.

The two-phase formation scenario is consistent with the
observations of quiescent galaxies at high redshifts $(z\sim 2)$
which are very massive ($\sim10^{11}~M_\odot)$ and compact $(R_e \sim 1$ kpc), usually called red
nuggets  \citep[e.g.][]{van2008confirmation,damjanov2009red,newman2010keck,whitaker2012large}. These objects
are a factor of 2-4 times smaller than current ellipticals  \citep[e.g.][]{daddi2005passively,trujillo2007strong,van2008confirmation}, and there is evidence
that they grow rapidly in size with little or no star formation 
 \citep[e.g.][]{van2010substantial,damjanov2011red,van20143d,buitrago2017cosmic}. It means that
whatever the growth process, it has the restriction to increase considerably
the size of galaxies without significantly increasing their star formation rate.
Indeed, the predominance of old stellar
populations in nearby massive ETGs is inconsistent
with large episodes of recent star formation, which suggests
that the mergers which form them either occurred
very long ago (at redshifts $z > 2$), or  they correspond to
 the coalescence of pre-existing old stellar populations \citep{thomas2005epochs,renzini2006stellar,graves2009dissecting}, suggesting that dry mergers 
could explain the growth of these objects \citep[e.g.][]{naab2009minor,trujillo2011dissecting,iodice2017fornax,mancillas2019probing}.  However, since merging processes are relatively rare in rich clusters due to their high
velocity dispersions, the late growth of ETGs must originate from a combination of other environmental mechanisms, such as tidal stripping, galaxy harassment, and dynamical friction around central galaxies \citep[e.g.][]{boselli2006environmental,cimatti2019introduction}. Galaxies can also experience significant pre-processing in filaments and groups before infalling into massive clusters \citep[e.g.][]{kuchner2022inventory}.

Some studies indicate that the environment plays an important role in this late evolution of galaxies. For instance, \cite{shankar2013size} suggest
that many compact ETGs at high redshifts have evolved into the
ETGs observed in high-density environments in the Local Universe. 
\cite{lani2013evidence} show that there is a strong relationship between the size
of quiescent galaxies and the environment where they are located, such that  massive objects
($M_\ast \geq 10^{11}~ M_\odot$) are significantly larger in
high-density environments. However, \cite{cypriano2006shrinking} find evidence for
the shrinking of cluster ellipticals at $z <0.08$, probably via tidal stripping, which is consistent with the result of \cite{matharu2019hst}, indicating that the disappearance of the compact cluster galaxies might be explained if, on average, $\sim 40$\% of them merge with their brightest cluster galaxies (BCGs) and $\sim 60$\% are tidally destroyed into the intracluster light  between $0 \leq z \leq 1$. 
Also, \cite{oogi2016mass} use numerical simulations
to investigate the role of dry mergers in the evolution of the mass-size relationship
of ETGs in massive halos and find that central galaxies in low-z clusters
with $M\sim 10^{14}~ M_\odot$ are potential
descendants of the red nuggets observed in $z \sim 2-3$. 
 \cite{yoon2017massive} find  that local early-type galaxies heavier
than $10^{11.2}~ M_\odot$ show a clear environmental dependence in mass–size relation, in such a way that galaxies are as
much as 20\%–40\% larger in the densest environments than in underdense environments.
On the other hand, some studies do not indicate a correlation between the environment and the size growth of galaxies
\citep[e.g.][]{allen2015differential,damjanov2015environment,saracco2017cluster}.

An important point about studying galaxies in clusters is that once inside these systems, the galaxies remain there, 
allowing  to connect high redshift and local objects
without needing to appeal to abundance matching to link
populations at different epochs \citep{guo2010galaxies,de2016morphological,wellons2017improved}.
Another point to be considered is that the
dynamics of galaxy clusters is closely associated with the orbits of member galaxies. Different Hubble types are usually related to  different orbits, with ETGs presenting more isotropic orbits than late-type galaxies \citep[e.g.][]{biviano2002eso,aguerri07,cava17,mamon2019structural}. At the same time, E and S0 galaxies 
show smaller velocity dispersions than spirals and irregulars \citep[e.g.][]{sodre1989kinematical, stein1997velocity, adami1998eso, biviano2002eso}. 
Also, \cite{nascimento2019influence}
find that the passive population in  systems with gaussian velocity distribution is the only family with lower velocity dispersion in massive clusters. \cite{morell2020classification} find a similar result showing that
ellipticals and lenticulars have the most isotropic orbits.

These examples of morphological and orbital segregation suggest that ETGs are former entrants into the potential of clusters. 
However the constant infall of galaxies into  clusters can produce a mixture between ancient and recent early-type objects.
According to \cite{berrier2008assembly},
the accretion times for $z = 0$ cluster members are quite extended, with $\sim$20\% incorporated into the cluster halo more than 
7 Gyr ago and $\sim$20\% within the last 2 Gyr. 
The effect of  infalling objects
can be studied using the distribution of the projected positions and velocities of cluster galaxies, 
due to the relationship between the region in the
projected phase space (PPS) and the time since infall of objects, as shown by the recent works of \cite{rhee2017phase}, \cite{pasquali19},  \cite{rhee2020yzics}, \cite{sampaio2021investigating}, and \cite{de2021roger}.
This connection makes galaxy clusters good laboratories to study environmental effects
on the growth of ETGs, since we can define galaxy
samples that inhabit the systems for different time scales.

In this work, we study the concentration index, radius, dark matter fraction, and color gradient of ETGs 
as a function of stellar mass and PPS regions
in regular massive clusters ($M_{200} \geq 10^{14}~ M_\odot$) at low redshifts. Since most of the growth of ETGs happened at $z \gtrsim 0.3$ \citep{lopez2012dominant,ownsworth2014minor},
we aim to understand the environmental influence on the late growth of ETGs 
according to the time since infall into their host clusters.
The paper is structured as follows: In Sect. 2 we present the data. The analysis is presented in Sect. 3. In Sect. 4 we give a summary of the results and our conclusions. Throughout this paper we adopt the cosmology with $\Omega_m = 0.3, \, \Omega_\Lambda=0.7,\, {\rm and}\, {\rm H}_0=100\,h\,{\rm km}^{-1}{\rm s}^{-1}{\rm Mpc}^{-1}$, with {\it h} set to 0.7. 


\section{DATA}

Our sample was obtained from the extended version of FoF group catalog originally identified by \citet{B06} and contains 5352 groups with 
$N > 5$ and  $0.03 \leq z \leq 0.11$, consisting of galaxies with  absolute magnitudes $M_r\leq -20.5$, and stellar masses in the range $10.4< \log{(M_\ast/M_\odot)} < 11.9$,\footnote{We  use  the notation “$\log{x}$” as indicating the decimal logarithm of $x$.}  with median $\sim 10^{11}~ M_\odot$. This  version is described in \citet{Lab10} and differs from the first one only in the area used (9380 square degrees from SDSS-DR7, compared to the original area of 3495 square degrees from the DR3).  We derived a refined central redshift by applying the gap technique \citep{A98, L07, L09} to the central (0.67 Mpc) galaxies. We also obtained a member list for each group using the ``shifting gapper" technique \citep{fadda1996observational,L09}, extending to $\sim 4$ Mpc around
the  group centers defined by \citet{Lab10}. 
The groups were then subject to the virial analysis, ana\-lo\-gous to that described in \citet{G98,P05,P07,Bi06,L09}. This procedure yields estimates of velocity dispersion ($\sigma_v$), radii ($R_{500}$, $R_{200}$) and masses ($M_{500}$, $M_{200}$) for most of the groups from the FoF sample. 
In the present work, we studied a subsample containing 107 massive clusters, with at least 20 galaxies within $R_{200}$, implying systems with $M_{200}\gtrsim 10^{14}{\rm M_\odot}$ extending to $2R_{200}$.

We added to the resulting sample  the stellar masses  from the \textit{galSpecExtra} table \citep{2004MNRAS.353..713K}, corresponding to the \textit{lgm\_tot\_p50}  parameter.
We also used the \cite{dominguez2018improving} catalog, which provides morphological T types for $\sim 670,000$ galaxies from SDSS, by training Convolutional Neural Networks (CNNs) with information from available sources such as Galaxy Zoo 2 \citep[GZ2,][]{2008MNRAS.389.1179L,2013MNRAS.435.2835W} and the catalog of visual classifications provided by \cite{nair2010catalog}, from which we included the probabilities of galaxies having a dominant bulge ($P_{bulge}$) and being S0 ($P_{S0}$). Our sample was also cross-matched with Korean Institute for Advanced Study Value-Added Galaxy Catalog \citep[KIAS VAGC,][]{parkchoi}, which provides some relevant photometric information, such as the rest frame color gradient $\Delta(g-i)$, and the concentration index $C=R_{90}/R_{50}$. Finally, we
added the S\'ersic indices $n$ in the $r$-band from the catalog of \cite{simard2011catalog}.



We defined a sample of 936  ETGs as objects with $T< 0$, $P_{bulge} > 0.5$ and $n> 2.5$ located in 48 clusters 
without substructures, according to the DS test \citep{dressler1988evidence};
and with Gaussian velocity distribution, according to the HD metric \citep[see][]{2013MNRAS.434..784R,decarvalho}. 
Both DS and HD indicators were used at the 90\% confidence level\footnote{The HD metric is not a statistical test  and so its confidence level must be understood as a confiability derived from bootstrap realizations, as described in \cite{decarvalho}.}, which means we restrict our study to systems likely to be virialized, 
avoiding the dynamical effects of interacting clusters and possible deviations from the expected homology of galaxy systems.

Galaxies are studied according to their loci in the PPS, which is built by normalizing the projected clustercentric distances and velocities by the virial radius, $R_{200}$, and velocity dispersion, $\sigma_v$, respectively.
To explore the PPS, we divide cluster galaxies into four subsamples, following the classification introduced by \cite{rhee2017phase}: ancient infallers (region E in their Figure 6), recent infallers (regions B \& C), intermediate infallers (region D), and first infallers (region A). These regions are related to the time since infall ($t_{inf}$) in Gyr: 

\begin{enumerate}
    \item[1)] Ancient $\rightarrow$   $6.5 \lesssim t_{inf} \lesssim 13.7$
    \item[2)] Intermediate $\rightarrow$ $3.6 \lesssim t_{inf} \lesssim 6.5$
    \item[3)] Recent $\rightarrow$ $0< t_{inf} \lesssim 3.63$
    \item[4)] First infallers $\rightarrow$ objects which are not completely fallen yet
\end{enumerate}

\begin{figure}
 \includegraphics[width=\columnwidth ]{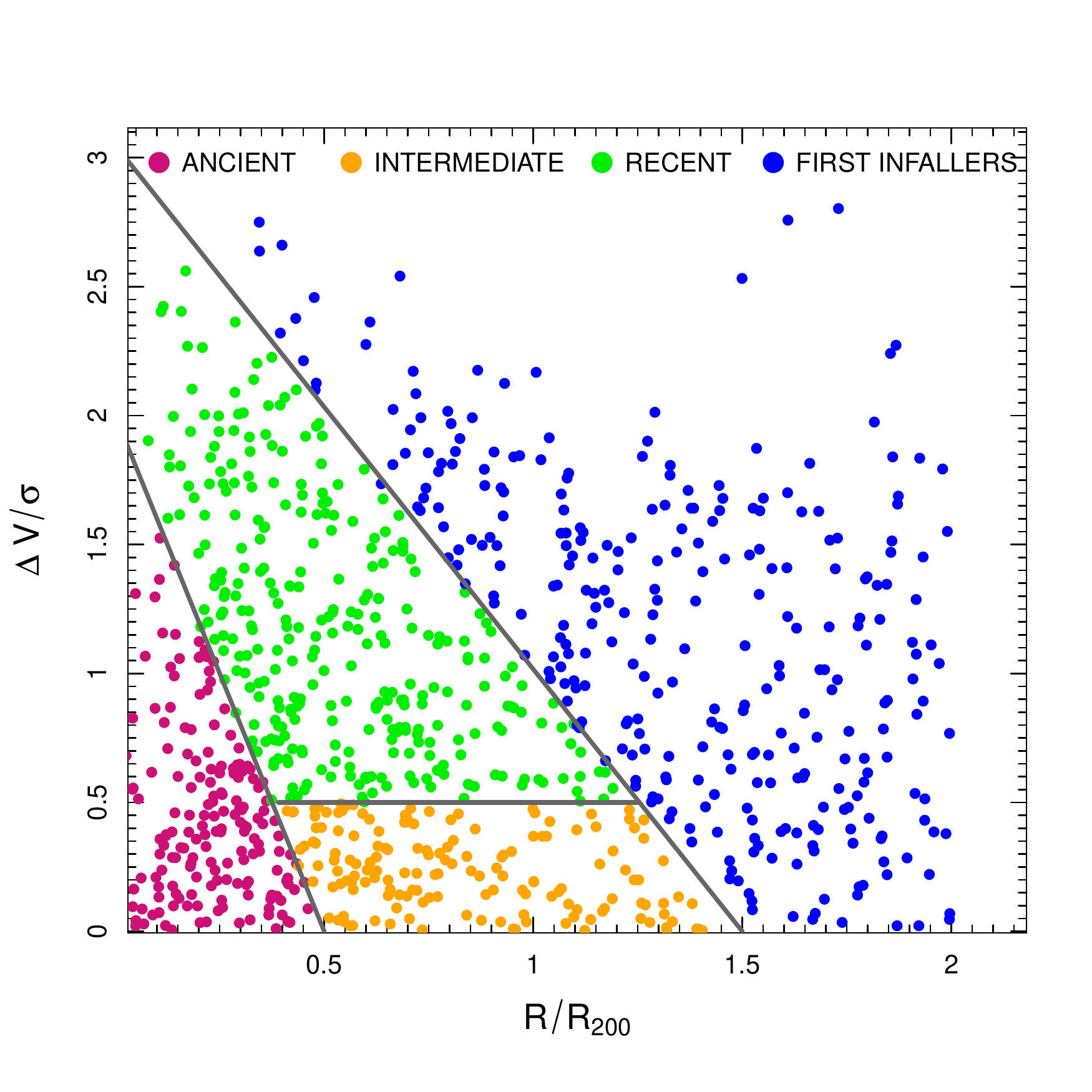}
 \vspace{-0.75cm}
 \caption{Distributions of ETGs in different PPS regions. Ancients objects are in red,
 intermediate in orange, recent in green, and the first infallers are in blue.}
 \label{fig1}
\end{figure}

In Figure \ref{fig1} we present the distribution of ETG galaxies in the PPS
regions of \cite{rhee2017phase}, following the approximation of \cite{song2018redshift}.
All the ancient (171 objects), intermediate (157 objects), recent (307 objects), and
first infallers (301 objects) are shown in this figure.

\section{Analysis}

In the following, we compare some properties of ETGs in the PPS samples,  also comparing them to a sample of isolated early-type galaxies taken from  the list of 1-member groups of
the catalog of \cite{yang2007galaxy}. This field sample, composed of 6670 objects, is defined with the same criteria (except membership) used to select the cluster ETGs.

\subsection{Concentration index}

\begin{figure}
\includegraphics[width=86mm,height=90mm]{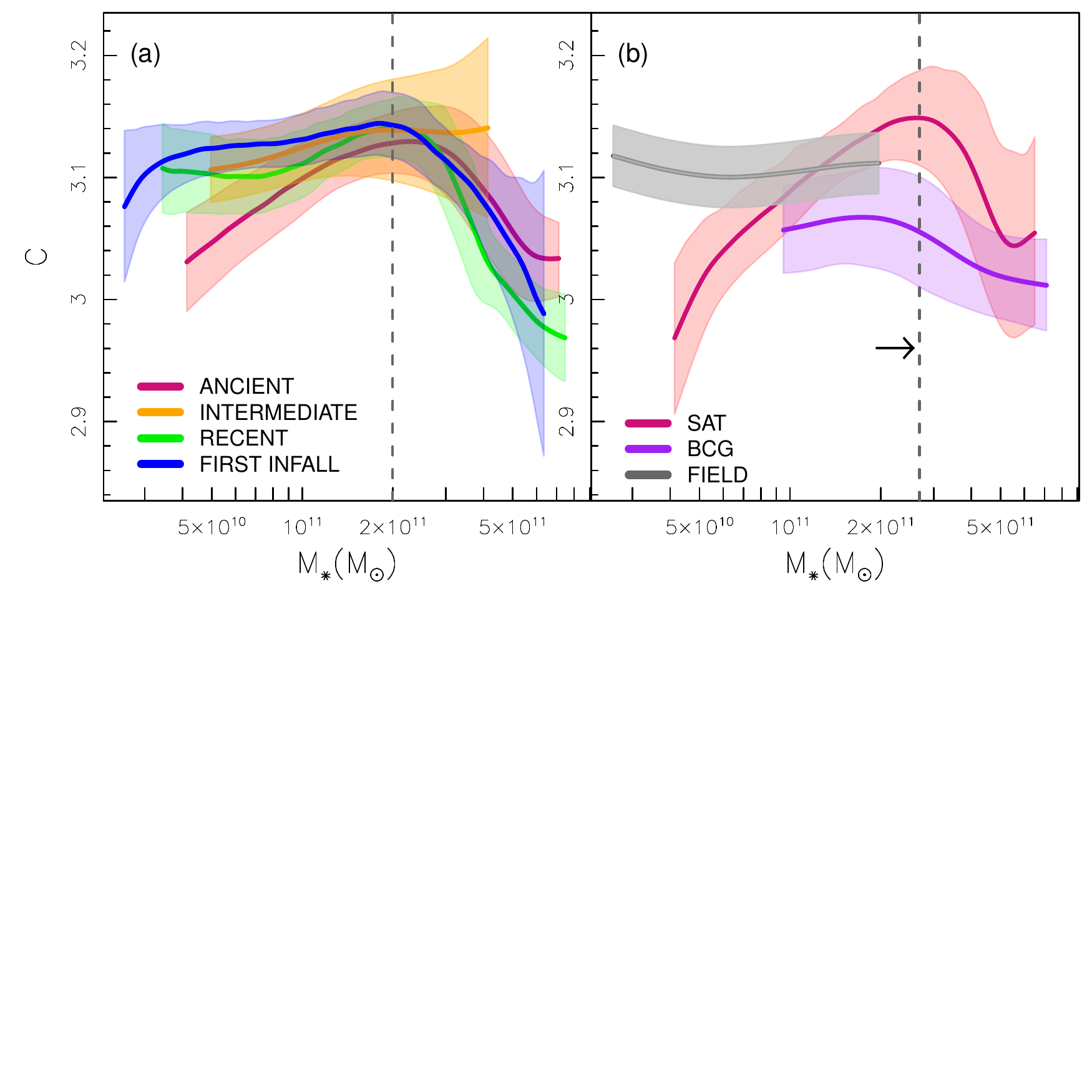}
\vspace{-4.5cm}
\caption{Stellar mass - concentration plane  of ETGs selected in the four PPS regions. (a) Ancients objects are in red,
intermediate in orange, recent in green, and the first infallers are in blue. (b) Satellites are in red, BCGs are in purple, and field galaxies are in gray.
The dashed vertical lines indicate
the behavior change around $2 \times 10^{11}~M_\odot$. Solid lines represent medians, while shaded areas show the confidence intervals calculated from 1000 bootstrap realizations. }
\label{fig2}
\end{figure}

A direct way to  study the late growth of ETGs is to consider them in the cluster environment and compare their
behavior in the stellar mass-concentration plane, taking into account their positions in the PPS, which is our basic aim in
this work. The concentration index reflects the current light distribution concerning $R_{50}$ and $R_{90}$, which can be
affected in different ways by the environment. Examining the concentration index as a function of the stellar mass, we find a significant change in the behaviour 
of $C$ around $M_{\ast} 
\approx 2 \times 10^{11} ~M_\odot \equiv \tilde{M}_{\ast}$, see panel (a) in Figure  \ref{fig2}. Objects less massive than  $\tilde{M}_{\ast}$ present a slight growth of $C$ with $M_{\ast}$, while more massive objects present  an opposite trend, except for those in the intermediate region.
\footnote{The intermediate region presents a small number of very massive galaxies (only 12 objects), a point to be discussed later. Also in this figure, we show the behavior of field galaxies.}

To better understand this result, we separated the sample of ancients, building a subsample of the brightest cluster galaxies (BCG) and another of satellites (SAT). The subsample of BCGs is  composed of the 48 most luminous ancient cluster ETGs (each taken from a cluster) having $M_r < -21.4$ and $M_\ast > 10^{11}~M_\odot$; and the SAT subsample is composed of the other 123 ancient galaxies. The behavior of these subsamples is shown in panel (b) of Figure \ref{fig2}. The behavior of SATs is the same (just slightly accentuated) as that of the total ancient sample; while BCGs have a narrower mass range (with no objects with $M_\ast < 10^{11}~M_\odot$) and  lower values of $C$  than SATs and field galaxies. We also should note that BCGs have a slightly decreasing behavior, while the curve of field galaxies is approximately flat. The flatter behavior of the field sample  suggests an environmental effect on the light distribution of cluster ETGs. For objects less massive than $\tilde{M}_{\ast}$ the PPS samples (especially the ancients) present a slightly higher growth (or a minor shrinkage) of $R_{90}$ with respect to $R_{50}$ (except for BCGs); and an opposite behavior  for objects with  $M_\ast > \tilde{M}_{\ast}.$

\begin{figure}
\includegraphics[width=86mm,height=90mm]{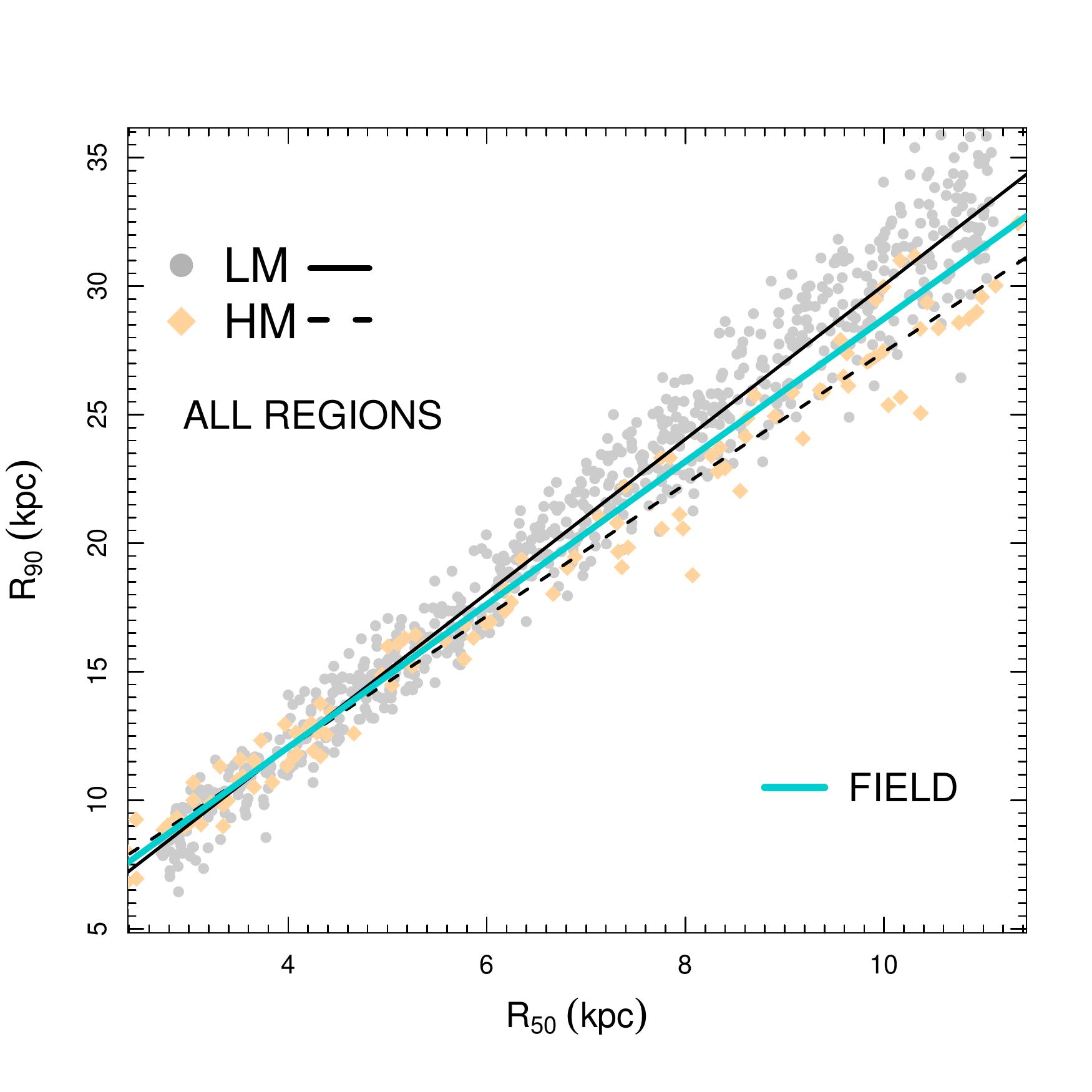}
\vspace{-0.55cm}
\caption{Regression lines for the $R_{50} \times R_{90}$ relations of  low mass (LM) and high mass (HM) ETGs taken from
all PPS regions. Gray circles indicate LM objects and beige diamonds indicate HM objects. The solid black line depicts LM ETGs while
the dashed black line depicts HM ETGs.  The solid cyan line is the regression line of the field ETGs.}
\label{fig3}
\end{figure}

According to \cite{rhee2017phase} there is a direct correlation between time since infall and tidal mass
loss, in the sense that objects longer in clusters, have experienced a higher rate of tidal mass loss.
A similar result is found by \cite{joshi2017preprocessing}.
Theoretically, tidal stripping
from the cluster potential tends to preferentially affect the outer
galaxy first, causing an  outside-in stripping  \citep[see][]{binney1987galactic}.
Similarly, tidal stripping from impulsive galaxy–galaxy
encounters (the harassment) preferentially affects the outer galaxy first \citep[see][]{smith2016preferential}. In addition, \cite{lokas2020tidal}
shows that, as a result of tidal stripping,  galaxies
weakly evolved (those with one pericentric passage) lose between 10\% and 80\% of their dark mass and less than 10\% of stars, while objects  strongly evolved (those with multiple pericentric passages) 
lose more than 70\% of dark mass and between 10\% and 55\% of stellar mass. All this suggests an effect on the light distribution of galaxies in their outermost parts, possibly implying a shrinkage of $R_{90}$ at different rates, being higher for objects more massive than  $\tilde{M}_{\ast}$, and lower for less massive objects. Running robust linear regressions between $R_{50}$ and $R_{90}$ for objects more or less massive than  $\tilde{M}_{\ast}$ (High Mass - HM and Low Mass - LM, respectively), we see that the relationship for LM objects is steeper -- see Figure \ref{fig3}. Also in this figure, field ETGs present an intermediate slope.
\footnote{We have not included the points of field ETGs so as not to saturate the figure.} To test the difference between slopes, we use
the analysis of variance (ANOVA) and find that the slopes $\beta_{LM}=3.03\pm 0.06$, $\beta_{HM}=2.69\pm 0.08$, and  $\beta_{FIELD}=2.78\pm 0.04$ are
significantly different at the 90\% confidence level. The different slopes imply small but significant differences between the median concentration indices for LM ($C=3.13\pm 0.03$) and HM ($C=3.03 \pm 0.05$) objects in all PPS regions.

The results of this section and the next ones  may be subject to two types of contaminants: (i) objects misclassified as cluster members, and (ii) contaminants with respect to the boundaries of the regions defined by \cite{rhee2017phase}. The first type of contaminants will be disregarded in this work since  our shift gapper + virial analysis code has been compared to a set of 24 galaxy-based cluster mass estimation techniques and proved to be among the best three \citep{old2015galaxy}, so the number of intruder galaxies remaining after applying the method should not be significant. The second type of contaminant is due to projection effects, i.e.,
the properties of galaxies identified in 3D differ from their properties when they are identified
on the PPS
\citep[e.g.][]{coenda2022reconstructing}. Indeed, \cite{rhee2017phase}  quantify the probability that
a galaxy in a particular location belongs to a particular infall region, and they also quantify the one-sigma
error on this probability due to cluster-to-cluster
variations, and differing lines-of-sight. To consider this, we assume the same contamination fractions in each region of the PPS as given by
\cite{rhee2017phase} (see their Figure 6). For each reference region, we randomly select points from the other three regions,
according to the contamination fractions, and reclassify the selected data points to the given reference region. After repeating the procedure 100 times, we check the variation of galaxy categorizations. The procedure allows checking how many galaxies were misclassified in a given class and by which of the other classes. According to \cite{rhee2017phase}  the overall highest numbers of contaminants come from the recent class, which significantly contaminates the others, especially the ancient class.
Indeed, mutual contamination between all regions appears in varying degrees. Still, our tests indicate that these unavoidable misclassifications, even in noteworthy cases such as the recent class contaminants, do not lead to significant changes in the trends, since they do not imply variations that exceed the confidence intervals in all studied bins. (see Appendix  \ref{A}).
 Another point possibly affecting our results is the different stellar mass intervals of the samples, as one can see in Figure \ref{fig2}.
 But running the analysis in the approximate same range  of stellar mass, $ 10.6\lesssim \log{(M_\ast/M_\odot}) \lesssim 11.4$, building subsamples of 857 (cluster) and 5821 (field) galaxies,  we find that the results are approximately the same. From now on, all  average properties of ETGs will be obtained for these reduced samples.

To further explore the structural  evolution of  cluster ETGs, in the following subsection we study  the behavior of $R_{50}$ and $R_{90}$ according to the
PPS regions.

\subsection{Growing (or not) in size}

\begin{figure}
\centering
\includegraphics[width=90mm,height=92mm]{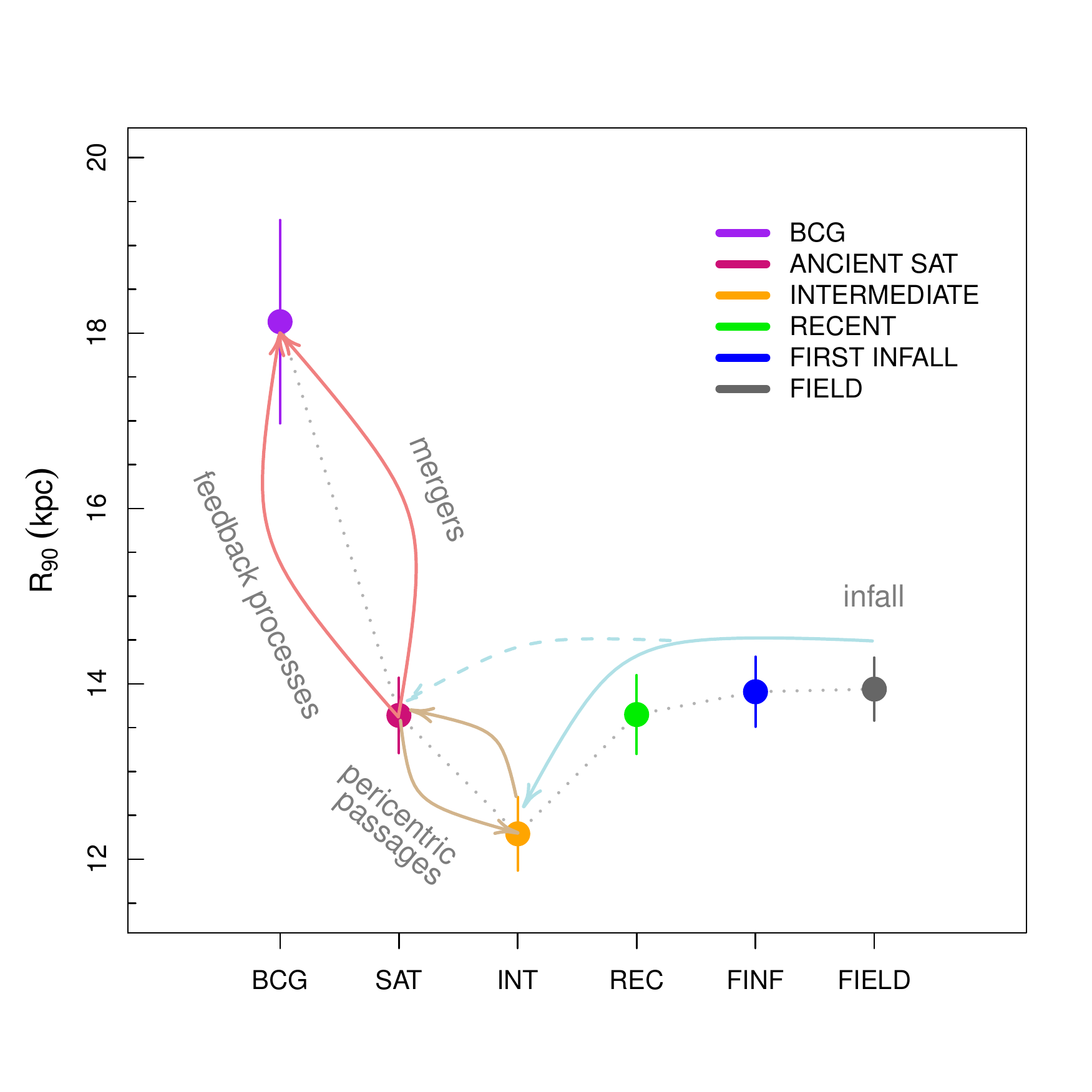}
\vspace{-0.8cm}
\caption{Evolution of $R_{90}$ (in kpc) across the PPS regions. 
Points and error bars indicate median values and bootstrap errors for 1000 resamplings. Ancients satellites are in red,
intermediate in orange, recent in green, first infallers in blue,  BCGs in purple, and field galaxies are in gray. Light blue arrows indicate the infall; orange arrows indicate the pericentric passages; and red arrows indicate processes leading to the  formation of the giant central ETGs.}
\label{fig4}
\end{figure}

The results presented so far suggest a differentiated growth (or shrinkage) of ETGs  as a function of their stellar mass. Now we analyze what happens to the radii $R_{50}$ and $R_{90}$ according to the PPS regions. 
First, we  study the variation of $R_{90}$ after the galaxy enters the cluster. As previously discussed, 
 $R_{90}$ is expected to be affected by tidal stripping and harassment during galaxy infall.
Figure \ref{fig4}  shows the evolution of this radius from the field to  BCGs. Note that the first stages of infall do not significantly change $R_{90}$, until the first pericentric passage occurs,  causing a remarkable shrinkage of the galaxies.
However, after multiple pericentric passages, galaxies grow in size again, reaching  values of $R_{90}$ $\sim$1.5 times larger than they had just after the first pericentric passage. This increase in size for ancient objects can be explained by the occurrence of mergers and/or feedback processes.
Mergers can increase the size of galaxies but usually also raise the S\'ersic index 
to $n> 5$ \citep[e.g.][]{hilz2013minor}. Although mergers are not frequent in massive clusters, dynamical friction acting on central galaxies can
reduce velocities and increase the rate of mergers and cannibalism on more massive objects \citep[e.g.][]{chandrasekhar1943dynamical,goto2005velocity,nipoti2017special,tamfal2021revisiting}.
In addition, objects that lose mass during the first pericentric passage can undergo dynamical self-friction, the process by which material that is stripped from
a subhalo torques its remaining bound remnant, which causes it to lose orbital angular momentum and favors mergers \citep{miller2020dynamical}.

\begin{figure}
\centering
\includegraphics[width=90mm,height=92mm]{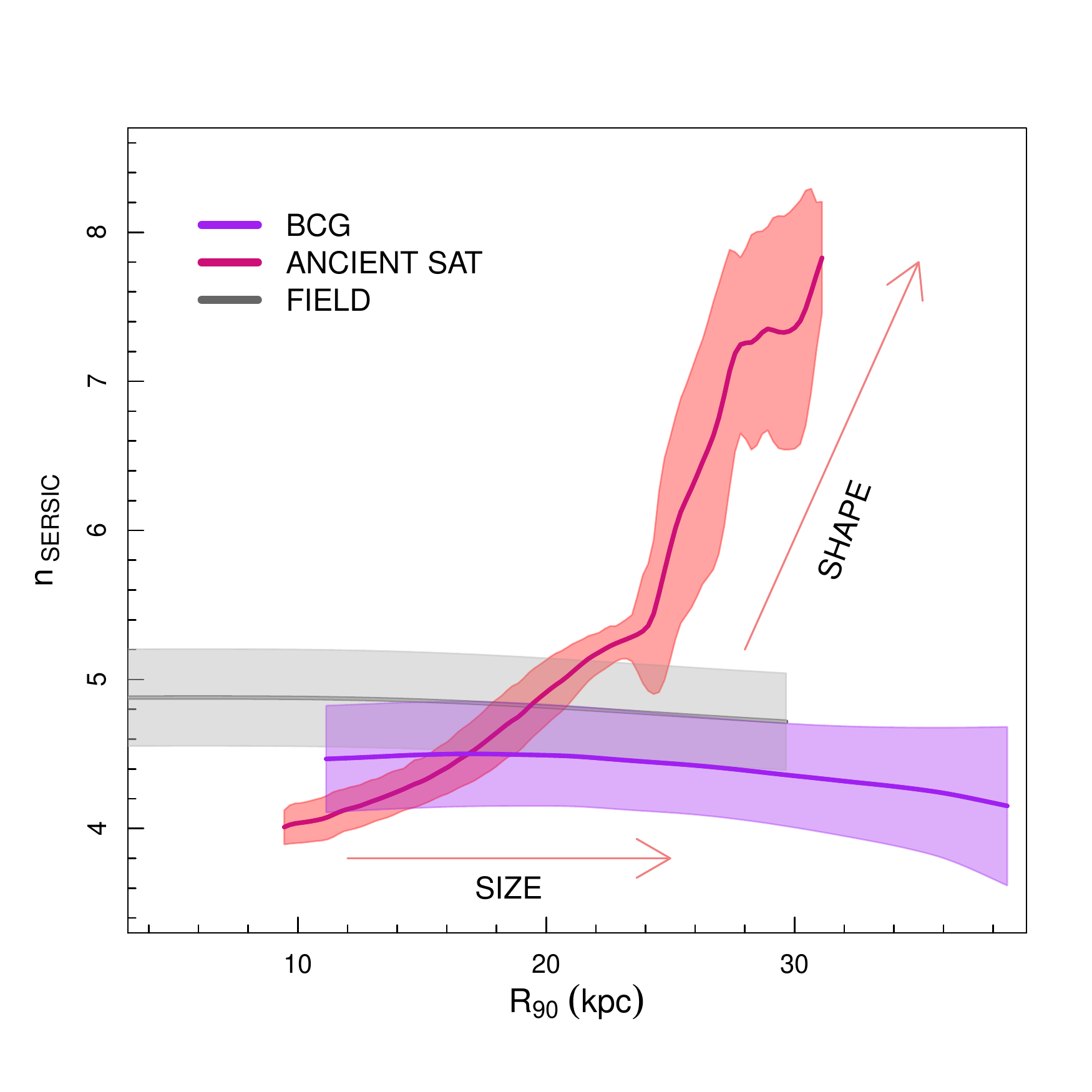}
\vspace{-0.8cm}
\caption{Relation between size given by $R_{90}$ and shape given by the S\'ersic index $n$. Red lines for ancient satellites, purple for BCGs, and gray for field ETGs. }
\label{fig5}
\end{figure}

On the other hand,  \cite{smith2016preferential} find that feedback effects, such as galactic winds or displacement of  black holes from the center of galaxies, can lead to an increase in size (via adiabatic expansion)  but not in the shape of central galaxies.
In Figure \ref{fig5}, we see the relation between
$R_{90}$ and the  S\'ersic index $n$ for ancient objects (satellites and BCGs) and field galaxies, as a comparison. Note in this figure that field galaxies and BCGs vary significantly in size but not in shape. At the same time, ancient satellites grow in both $R_{90}$ and $n$, with the S\'ersic index reaching $n\approx 8$, which could be obtained after two intermediate
(mass ratio 1:5) dry mergers \citep{smith2016preferential}. This result indicates that BCGs grow preferentially via feedback mechanisms, in agreement with \cite{ascaso2010evolution} that 
explain the increase in size and the non-evolution in the S\'ersic shape parameter of the BCGs in the last 6 Gyr through 
feedback processes. Alternatively,
BCGs may also have reached their final size in earlier stages of cluster formation.  Assuming that part of the ancient satellites can merge with BCGs, the result indicates that the latter can grow via both mechanisms.

\begin{figure}
\centering
\includegraphics[width=90mm,height=92mm]{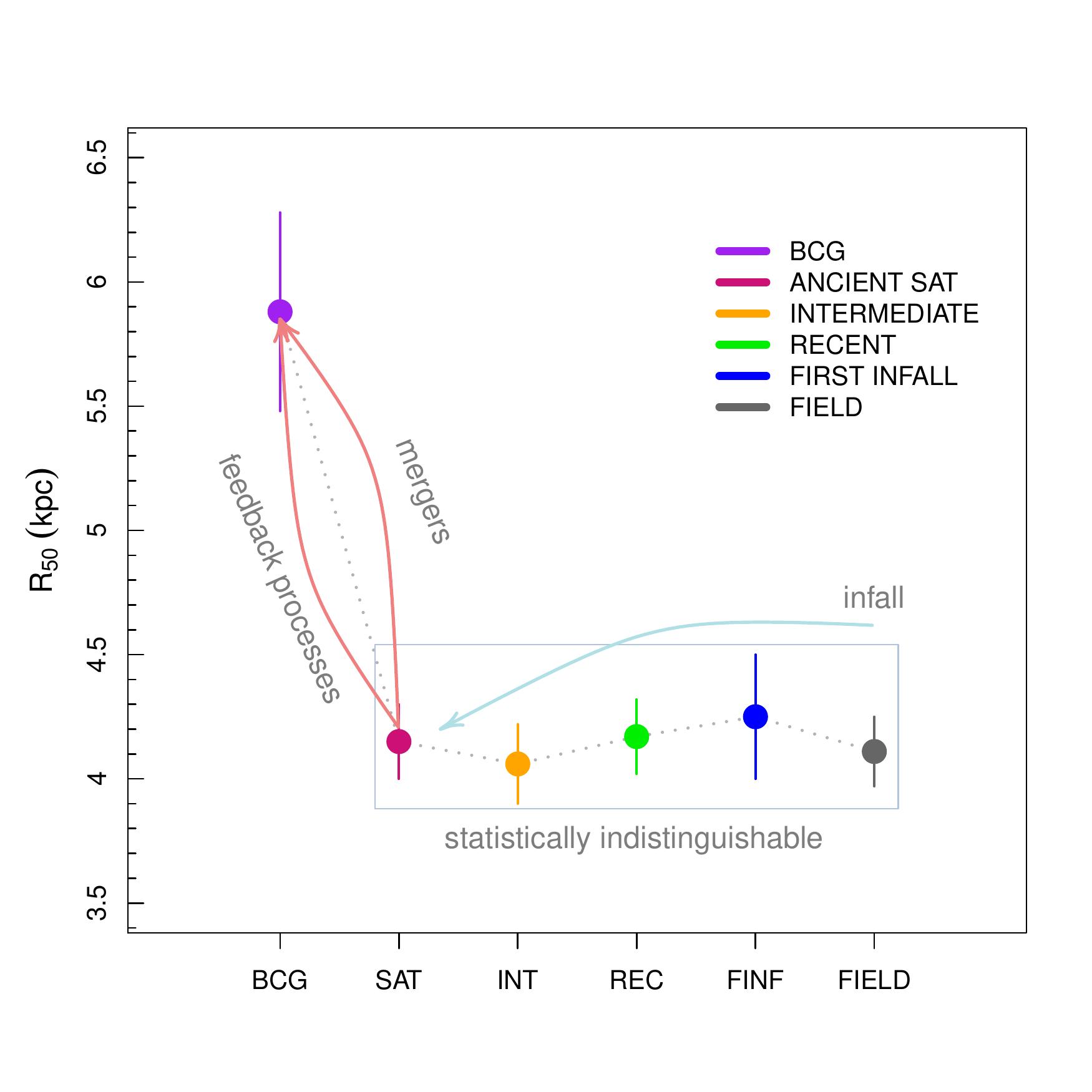}
\vspace{-0.8cm}
\caption{Evolution of $R_{50}$ (in kpc) across the PPS regions. 
Points and error bars indicate median values and bootstrap errors for 1000 resamplings. Ancients satellites are in red,
intermediate in orange, recent in green, first infallers in blue,  BCGs in purple, and field galaxies are in gray. Light blue arrows indicate the infall; orange arrows indicate the pericentric passages; and red arrows indicate processes leading to the  formation of the giant central ETGs. }
\label{fig6}
\end{figure}

To complete the analysis, we finally show how $R_{50}$ varies according to the PPS regions. In Figure \ref{fig6} we see that the values of $R_{50}$ do not change significantly during the infall, indicating that environmental effects do not change the light distribution at this scale. However, for ancient objects, we notice a remarkable growth in size up to the BCG values.

\subsection{Color gradient}

Several studies show that local  ETGs have negative color gradients, indicating
that their stellar populations become
bluer towards the galaxy outskirts \citep[e.g.][]{peletier1990ccd, la2009origin, gargiulo2012spatially}. This behaviour also supports the scenario
where ETGs  assemble most of their stellar mass at high redshifts
$(z \gtrsim 2)$, and then experience passive evolution interspersed by  dry minor mergers over cosmic time \citep[e.g.][]{di2009survival,hilz2013minor}.
The study of ETGs through the PPS regions of clusters (i.e., their time evolution within a dense environment)
can help us to better understand the late environmental effects on ETGs. In this work,
the rest frame color gradient $\Delta(g-i)$
is defined as the
difference in color of the region with $R < R_{pet}$
from that of the annulus with $0.5 R_{pet} < R < R_{pet}$,
where $R_{pet}$ is the Petrosian radius. According to this definition, a negative
color difference means a bluer outside (or redder centers) \citep{parkchoi}.

\begin{figure}
\includegraphics[width=86mm,height=90mm]{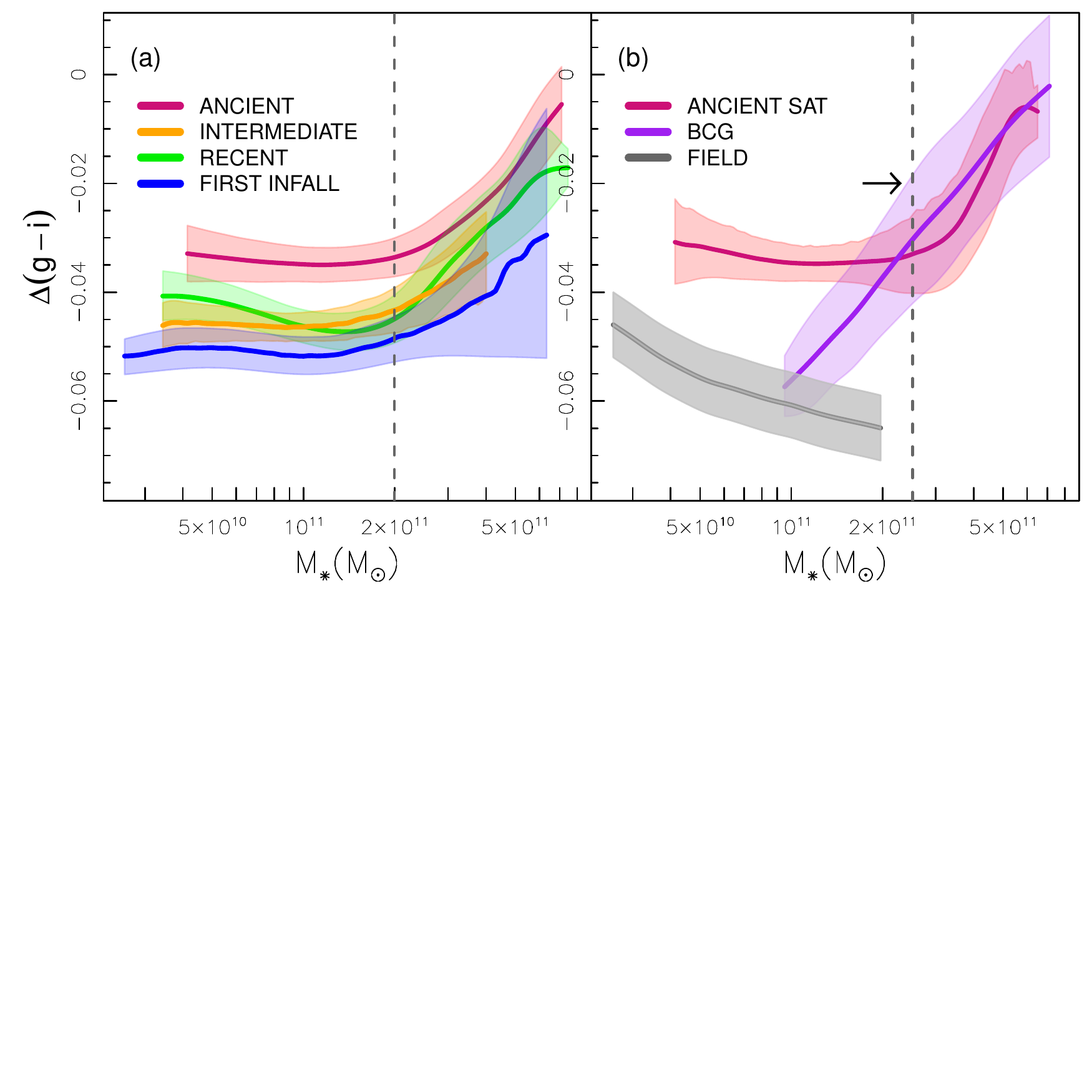}
\vspace{-4.55cm}
\caption{Color gradient dependence on the stellar mass  of ETGs selected in the four PPS regions. Ancients objects are in red,
intermediate in orange, recent in green, and the first infallers are in blue. The dashed vertical line indicates
the behavior change around $\tilde{M}_{\ast}$. Solid lines represent medians, while shaded areas show the confidence intervals calculated from 1000 bootstrap realizations.}
\label{fig7}
\end{figure}

In Figure \ref{fig7}, we see how color gradients $\Delta (g-i)$ vary with the stellar mass of the ETGs. The first point to note in panel (a) of this figure is  a change in the general behavior at approximately the same mass where we observe a transition in the concentration index, $\tilde{M}_{\ast}$. 
For lower masses, color gradients are negative and approximately constant in all samples, while for  $M_{\ast}>\tilde{M}_{\ast}$ color gradients tend to become less negative with $M_\ast$. Two-sample permutation comparisons indicate (at the 95\% C.L.) that the ancient population has the most positive color gradients ($-0.034\pm 0.004$), while the recent $+$ intermediate samples have in-between values ($-0.045\pm 0.004$), and the first infallers are the ones with the most negative color gradients ($-0.051\pm 0.003$). 
This result points to objects longer in the clusters having less negative color gradients. At the same time,  more massive objects show an increasing trend of the color gradient with $M_\ast$ in all PPS regions.
Usually, the centers of ETGs are older and more metal-enriched than their outskirts, producing  negative color gradients. To make them less negative, the action of
tidal stripping could be removing the outskirts and/or dry mergers would be flattening the color gradients. According to  the relationship between infall time and tidal stripping,  ETGs with a stellar mass smaller than $\tilde{M}_{\ast}$ present a small shrinkage of $R_{90}$ (or smaller sizes with respect to $R_{50}$ -- as we found in the previous sections)
indicating that tidal stripping could explain the behavior of color gradients for these objects. It is important to note that this behavior does not change with stellar mass, as opposed to what happens with more massive objects, as we can see in panel (b) of
Figure  \ref{fig7}, for which the color gradients increase almost linearly with stellar mass, indicates that mergers may affect these objects by reducing (approaching zero) the mean colour gradients. The flattening can also be favored by preprocessing, where more massive objects can pass through mergers in groups before entering the clusters.
As for field objects, low-mass ETGs have color gradients similar to those of first infallers. The values become more negative for objects with higher stellar masses, possibly indicating the absence of mechanisms pruning the blue edges of these objects.

\subsection{$M_\ast - \sigma$ relation}

The scaling between stellar mass and velocity dispersion is one of the fundamental connections observed in ETGs\citep[e.g.][] {shen2003size,hyde2009luminosity,roy2018evolution,zahid2018stellar,tortora2018last},
and it directly follows   the Faber-Jackson relationship \citep{faber1976velocity}. Stellar
velocity dispersion depends on the gravitational potential and therefore relates galaxies to their dark matter halos
\citep[e.g.][]{evrard2008virial,schechter2016new}. At the same time, the stellar mass indicates the amount of baryonic mass converted into stars. \cite{zahid2018stellar} use data from the Illustris simulations to show that the stellar velocity dispersion is equal to the velocity dispersion in the dark matter halo. 
The observed relationship follows $\sigma \propto M_{\ast}^{0.3}$ for $M_{\ast} > 10^{10.3} M_\odot$
 \citep[e.g.][]{cappellari2016structure,zahid2016scaling}, 
which is approximately  the same relation between  $\sigma_{DM}$ and  $M_{200}$ for dark matter particles \citep{zahid2018stellar}. Although the $M_\ast-\sigma$  relation does not show significant evolution at $z < 0.7$ \citep{zahid2016scaling},  environmental effects may be affecting the relation for clusters ETGs with $M_\ast > 10^{11.2} M_\odot$, since these objects experience  major and
minor mergers more frequently in high-density environments \citep{yoon2017massive}. 
Indeed, the results presented in the
previous sections indicate that dynamical friction
may slow down ancient satellites that vary greatly in size and shape, suggesting mergers in clusters or possibly in infalling groups where preprocessing took place. On the other hand,
less massive ETGs  seem to be affected by tidal stripping in the cluster environment.

 We study the behaviour  between  $M_\ast$ and  $\sigma$ for galaxies in the PPS subsamples.  
The central stellar velocity dispersions are aperture corrected by the equation of  \cite{cappellari2006sauron}:

\begin{equation}
\sigma = \sigma_{\rm{fiber}} \left(8 r_{\rm{fiber}}\over R_e \right)^{0.066}
\label{eq1}
\end{equation}

\noindent where $\sigma_{\rm{fiber}}$ is the estimated velocity dispersion in SDSS,  $R_e$ is
effective or the angular half-light radius in arcseconds, and  $r_{\rm{fiber}}$ is the
radius of SDSS fibers, $1.5^{ \prime \prime}$. We only use ETGs with $100 < \sigma < 420 ~{\rm km~s^{-1}}$, since values below 100 ${\rm km~s^{-1}}$ are unreliable \citep[e.g.][]{bernardi2003early}, and 
 SDSS uses template spectra convolved to a
maximum velocity dispersion of 420 ${\rm km~s^{-1}}$ \citep[see e.g.][]{yoon2020dependence}. The effective radii are calculated by 

\begin{equation}
R_e = a_{\rm deV}\sqrt{b/a}
\label{eq2}
\end{equation}

\noindent where $a_{\rm deV}$ and $b/a$ are the semimajor axis length and the axis
ratio from the de Vaucouleurs fit, respectively. We only use ETGs with $b/a > 0.3$ to avoid
edge-on objects. The restrictions on $\sigma$ and $b/a$ reduce the sample to 912 galaxies.
The parameters  $a_{\rm deV}$, $b/a$, and $\sigma_{\rm fiber}$  in $r$-band
are taken from the catalogs PhotObjAll and SpecObjAll of DR15
\citep{aguado2019fifteenth}.

\begin{figure*}
\includegraphics[width=160mm]{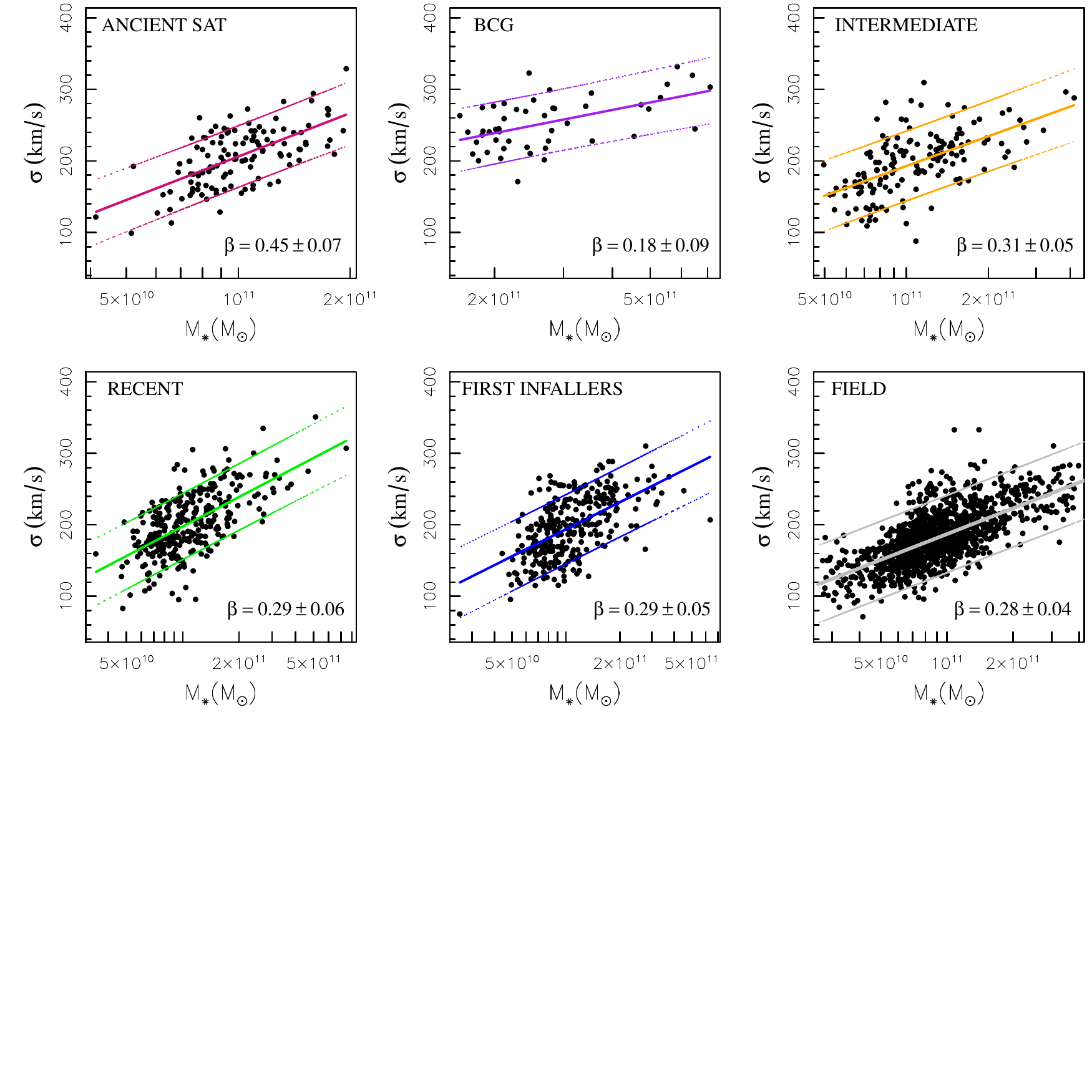}
\vspace{-5.25cm}
\caption{Stellar mass versus central velocity dispersion of ETGs selected in the four PPS regions. Linear fits plus 1$\sigma$ errors for ancient satellites are depicted in red lines, with BCGs in purple,
intermediate in orange, recent in green,  first infallers  in blue, and field galaxies in gray lines.}
\label{fig8}
\end{figure*}

In Figure \ref{fig8}, we see all the robust linear regressions  between central velocity dispersion and stellar mass  of ETGs in the  PPS regions (separating the ancients into BCGs and satellites) plus the field ETGs. Using ANOVA to compare the slopes from each sample we find that most of the samples present a similar slope, $\beta \approx 0.3$, except with respect to
the ancient objects, for which we find $\beta = 0.45 \pm 0.07$ for the
satellites, and $\beta = 0.18 \pm 0.08$ for the BCGs, statistically distinct values at the 95\% confidence level.

The slopes indicate how much change there is in the central velocity dispersion when the stellar mass changes, with slopes significantly above or below $\approx 0.3$ indicating a higher or lower fraction of dark matter in galaxies.
\cite{zahid2018stellar} showed that the central stellar velocity dispersion of quiescent galaxies is proportional to the dark matter halo velocity dispersion. Therefore, our results indicate an increase of  dark matter in ancient satellites and a deficit in BCGs. This is consistent with what we find in the previous sections, i.e., satellites in the central region of clusters are probably going through mergers (see, for example, Figure \ref{fig5}). 
Although mergers do not occur in clusters with high rates at low-z,  there are several observational evidence for dry merging at $z < 0.3$ in central galaxies of groups and clusters \citep[e.g.][]{mcintosh2008ongoing,liu2009major,rasmussen2010witnessing,edwards2012close}. In particular, \cite{liu2015ongoing}  have obtained a major dry merger rate of $0.55\pm 0.27$ merger per Gyr at $z \sim 0.43$.
Even such a low merger rate can promote the growth of dark matter halos in ancient ETGs since they have  time since infall $\geq$ 6.5 Gyr.
On the other hand, the constancy of the S\'ersic index does not indicate significant mergers for the BCGs, and the shallower slope in the $M_\ast-\sigma$ relation must be related to feedback mechanisms and/or to the fact that the BCGs could not constitute virialized configurations. The presence of self-interacting DM in haloes can also explain
lower  central DM densities  by making the core radius larger \citep[e.g.][]{rocha2013cosmological,di2017rumble}. Lower DM fractions could also be associated with diffuse DM haloes \citep{alabi2017sluggs}, or due to 
non-universal IMF and non-homology of ETGs  \citep[see e.g.][]{tortora2022central}.

\subsection{Difference
between dynamical and stellar mass}

Another way to explore the distribution of dark matter in ETGs through the PPS regions is to consider the difference between the  dynamical (virial) mass and the stellar mass of galaxies. We can estimate the dynamical mass using the 
following expression:

\begin{equation}
M_{\rm dyn} \simeq K(n) {R_e \sigma^2\over G}
\label{eq3}
\end{equation}

\noindent \citep[see][]{poveda1958masses,nigoche2019quantity}, where $K(n)$ is a scale factor 
that depends on the Sérsic index
$n$ as follows

\begin{equation}
K(n) = 8.87 -0.831 n + 0.00241 n^2
\label{eq4}
\end{equation}

\begin{figure}
\includegraphics[width=\columnwidth]{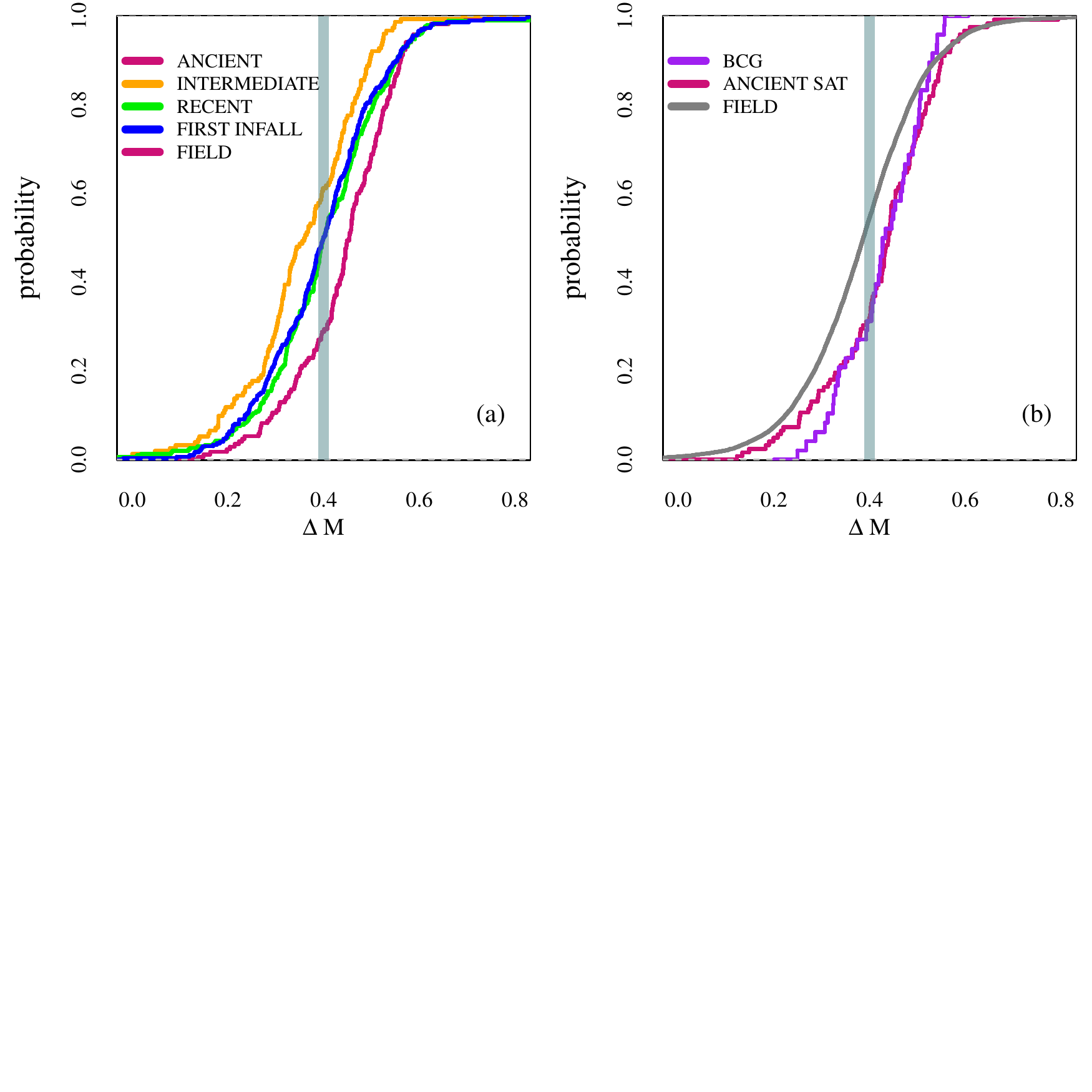}
\vspace{-4.75cm}
\caption{Cumulative distribution functions of the logarithmic difference
between dynamical and stellar mass, $\Delta M$. In panel (a) we show ancients objects in red,
intermediate in orange, recent in green, and the first infallers in blue. In panel (b), we show BCGs in purple,
ancient satellites in red, and field ETGs in gray.
The shaded area in both panels indicates the
range of median values of $\Delta M$ for the samples with $z<0.12$ studied by  \protect \cite{nigoche2019quantity}.}
\label{fig9}
\end{figure}

\noindent \citep{cappellari2006sauron}, with the
S\'ersic indices  taken from the catalog of \cite{simard2011catalog}. We also define
the ratio $M_{\rm dyn}/M_\ast$ as

\begin{equation}
\Delta M\equiv  \log{(M_{\rm dyn}/M_\odot)} -  \log{(M_\ast/M_\odot)}
\label{eq5}
\end{equation}

\noindent following the work of \cite{nigoche2019quantity}. We present the cumulative distribution functions of $\Delta M$ for the PPS samples in Figure \ref{fig9}, and use the Conover test to compare them against each other
\citep{conover1979multiple}.  We use the version  from
the DescTools R package \citep{signorell2016desctools}. The test
performs a multiple comparison between the datasets and verifies whether the cumulative distribution function (CDF)
of one does not cross the CDF of the other at the 90\% C.L.\footnote{See a previous astrophysics application of this test in  \cite{morell2020classification}.}
The test indicates significant differences between ancient objects and all other samples, as well as intermediate objects and all others, not indicating a significant difference between recent, first infallers, and field ETGs. We also find no significant difference between the BCGs and the ancient satellites.
In Figure \ref{fig9}, we show the cumulative distribution functions of the logarithmic difference
between dynamical and stellar mass for the PPS samples in panel (a), and BCG, SAT, and field samples in panel (b). 
The shaded areas in both panels indicate the
range of median values of $\Delta M$ for the samples with $z<0.12$ studied by \cite{nigoche2019quantity}, 
[0.388,0.411] for low-z ETGs. In comparison, our PPS samples have the following median values:
$0.431\pm 0.015$ (all ancient), $0.358\pm 0.028$ (intermediate), $0.399\pm 0.013$ (recent), and $0.398\pm 0.017$ (first infallers). Separating the ancients: $0.421\pm 0.019$ (BCGs) and $0.438\pm 0.016$ (satellites), while the field ETGs present $0.389\pm 0.011$. The medians  and respective errors are computed from 1000 bootstrapped resamplings. This result suggests that ancient ETGs have a higher fraction of dark matter, while intermediate objects have a lower fraction, with the recent, first infall, and field subsamples having values in the middle. Not only does this reinforces the idea of environmental effects on the distribution of dark matter in cluster ETGs, it also gives us clues  to understand the process taking into account their orbits and trajectories through the PPS of the clusters.
In fact, the result seems to indicate that objects with an infall time $\lesssim$ 3.63 Gyr are not significantly affected by dynamical processes that affect the dark matter fraction ($f_{DM}$), which remains at the level of $\sim$60\% (after translating to linear scale), agreeing with the median $f_{DM}$
 found by \cite{nigoche2019quantity} for low-$z$ samples. At the same time, ancient ETGs have a slight (but significant) increase in $f_{DM}$ to 65\%, while intermediate ETGs have $f_{DM}$ decreased to 56\%. 
 
 The decrease in $f_{DM}$ during the intermediate phase could be the result that objects in this region 
 had their first pericentric passage recently and may have lost part of their mass at this stage.  Indeed, \cite{gill2005evolution} report that cluster galaxies with low projected velocities, usually identified as backsplash galaxies, lose a fraction of their mass in their first passage through the cluster core. \cite{muriel2014galaxy}
find that ETGs with very low projected velocities $(|\Delta V| < 0.5 \sigma_{cl}$), where $\sigma_{cl}$ is the velocity dispersion of the host cluster, are systematically less massive than ETGs at higher velocities. To compare,
our intermediate population typically has $|\Delta V|\approx 0.25 \sigma_{cl}$. 
In addition, \cite{rhee2017phase} find a significant tidal mass loss of dark matter after the first pericenter passage; a similar result is found by \cite{joshi2017preprocessing}.
Interestingly, our intermediate population  presents a cutoff in stellar mass around $10^{11.6}~M_\odot$, while the other populations reach $\sim 10^{11.9}~M_\odot$. This reinforces the conclusion that  ETGs in this region of the PPS have either lost part of their mass (both stellar and dark matter) and/or are unable to add mass through dynamical processes. 
Finally,  we need to reconcile the higher $f_{DM}$ in ancient ETGs with the DM deficiency obtained from the $M_\ast-\sigma$ relation for BCGs. It is well known that BCGs have larger $R_e$, at fixed stellar mass than the general early type
population and their velocity dispersion increases less with
stellar mass \citep[e.g.][]{bernardi2009evolution}.
From Equations \ref{eq1} and \ref{eq3} we see that larger $R_e$ decreases $\sigma$ and increase $M_{\rm dyn}$, which could partially explain the result. The increase in $R_e$ can be associated with feedback effects and other effects we mentioned in the previous section. Understanding the physical processes driving the evolution of BCGs is essential for studying the formation history of galaxy clusters. Our analysis indicates
early-type BCGs differ significantly from the general ETG population in low-z clusters. Still, a more specific study on the evolution of BCGs will be presented in a forthcoming paper.

\section{Discussion}

The study of cluster ETGs is essential to improve our understanding of the structure
formation and the galaxy–environment connection. 
Advances in this topic may allow us to understand  how the environment may or may not favor the late evolution of ETGs. The size growth of ETGs is a long-term process
that can last until the present. \cite{andreon2018cosmic} shows that the environment has been affecting the sizes of ETGs since
at least $z\approx 2$. Merging seems to be the main contributor to the size evolution of ETGs at $z \lesssim 1$ \citep[e.g.][]{lopez2012dominant}, with
 massive local galaxies assembling $\sim 75$\% of their total stellar mass at $0.3<z<3.0$  \citep{ownsworth2014minor}, indicating 
 the need for late growth of ETGs at $z<0.3$. 
 However, the growth in size and mass of cluster ETGs is disfavoured by tidal effects, apart from the fact that mergers are less efficient in massive clusters.
 In this context, a significant result is found by 
 \cite{rhee2017phase}, showing a direct correlation between time since infall and tidal mass
loss, in the sense that objects longer in clusters have experienced a higher rate of tidal mass loss. In addition, \cite{matteuzzi2022newcomers} show that the evolution in size and mass of cluster ETGs is related to the acquisition of new galaxies by the clusters and the transformation of member galaxies located at large clustercentric distances at $z=0.85$, which end up being
massive satellite ETGs at $z=0$. Also, \cite{tran2008late} study BCGs and their early-type companions at $z\sim 0.37$ as an example
of late stellar assembly of massive cluster galaxies via major merging.

 In the present work, we present a study of cluster ETGs distributed in PPS regions defined by \cite{rhee2017phase} and \cite{song2018redshift}. The samples containing the oldest objects
 (ancient and intermediate ETGs)  have the time since infall  $\gtrsim 3.6$ Gyr,  which places their entry into the cluster at $z \gtrsim 0.32$. For the recent and first infallers ETGs, their time since infall is $\lesssim 3.6$ Gyr.
 Despite this difference in $t_{inf}$, all the cumulative galaxy changes, before and after the infall, are roughly called late (or residual) growth since we define  all the ETGs using the same criteria. So the differences we find between the PPS regions are, in principle, thought as a result of environmental effects in the clusters.
 
A possible caveat to the whole analysis is that the sample of ETGs has 77\% of ellipticals (E) and only 23\% of lenticulars (S0). We separate E from S0 following \cite{barchi2020machine}, considering ellipticals all objects with $T < -2$, and lenticulars
all those with  $-2 \leq T  < 0$ with  $P_{S0} > 0.6$.
Removing the S0, we  reproduce all the results presented in the previous sections. But using only the S0, the results become statistically inconclusive.
 Another difficulty inherent to this work is associated with the mutual contamination of objects from different regions of the PPS, as discussed in Section 3.1 and Appendix \ref{A}.

Despite these limitations, the present work achieves some significant results capable of shedding light on the problem of late growth of ETGs in galaxy clusters.
Our main findings are:

\begin{enumerate}
\setlength\itemsep{1em}
    \item We find a significant change in the behaviour of the concentration index $C$ around $\tilde{M}_{\ast}$, see Figure  \ref{fig2}. Objects less massive than this present a slight growth of $C$ with $M_{\ast}$, while more massive objects present  an opposite trend.  The flatter behavior of
the field sample, especially regarding ancient satellites, suggests an environmental effect on the
the light distribution of cluster ETGs. 

\item HM objects in all PPS regions show less growth (or more shrinkage) of $R_{90}$ concerning $R_{50}$ than LM objects
(Figure  \ref{fig3}), and the infall seems to decrease $R_{90}$ and not affect $R_{50}$ (see Figures  \ref{fig4} and  \ref{fig6}). But in the transition from  the intermediate to the ancient region, we find a remarkable increase of $R_{90}$ and $R_{50}$.

\item We also find that field galaxies and BCGs vary significantly in size but not in shape, while ancient satellites grow in both $R_{90}$ and $n$, with the S\'ersic index reaching $n\approx 8$, indicating that BCGs grow preferentially via feedback mechanisms at low-z (or that they have already reached their final size in earlier stages of cluster formation) while central satellites probably experience a growth stage via dry mergers.

\item For LM objects, color gradients are negative and approximately constant in all samples, while for  HM ETGs color gradients tend to become less negative with $M_\ast$. The ancient population has the most positive color gradients ($-0.034\pm 0.004$), while the recent $+$ intermediate samples have in-between values ($-0.045\pm 0.004$), and the first infallers are the ones with the most negative color gradients ($-0.051\pm 0.003$).
This result points to objects longer in the clusters having less negative color gradients.

\item From the study of the $M_\ast-\sigma$ relation, we find an increase of dark matter in ancient satellites and a deficit
in BCGs, indicating mergers for satellites and feedback mechanisms (or a more complex structure plus previous evolution) for BCGs.

\item  Estimating the dynamical mass of galaxies, we find that ancient ETGs have a higher fraction of dark matter, while intermediate objects have a lower fraction, with the recent and first infall samples having values in the middle.

\item Finally, results (v) and (vi) can be reconciled for BCGs if mechanisms capable of increasing $R_e$ act on these objects.

\end{enumerate}

These results favor a scenario where cluster ETGs  experience environmental influence the longer they remain and the deeper into the gravitational potential they lie. 
Our findings indicate a combination of  tidal stripping + harassment, which predominate during infall, followed by mergers + feedback effects affecting ancient satellites and BCGs, respectively.
 The competition between tidal stripping and mergers is probably the key element for  understanding the late evolution of ETGs toward their observed properties. At the same time, we should assume that these processes are superimposed on  pre-processing mechanisms that may have happened in galaxy groups before infall.
 A more detailed study of this scenario is necessary, constituting a point to be 
 further investigated in a future work.

\section*{Acknowledgements}
We thank the anonymous referee for  very helpful suggestions.
ALBR thanks the support of CNPq, grant 316317/2021-7 and FAPESB INFRA PIE 0013/2016. RSN thanks the financial support from CNPq, grant 301132/2020-8.
PAAL thanks the support of CNPq, grants 433938/2018-8 e 312460/2021-0. CCD thanks the support by the Coordena\c c\~ao de Aperfei\c coamento de Pessoal de N\'{\i}vel Superior - Brasil (CAPES) - Finance Code 001, the Programa Institucional de Internacionaliza\c c\~ao (PrInt - CAPES), and the Brazilian Space Agency (AEB) for the funding (PO 20VB.0009). MHSF thanks the financial support by the Coordenação de Aperfeiçoamento de Pessoal de Nível Superior - Brasil (CAPES) - Finance Code 001.

This research has made use of the SAO/NASA Astrophysics
Data System, the NASA/IPAC Extragalactic Database (NED) and
the ESA Sky tool (sky.esa.int/). Funding for the SDSS and SDSS-II
was provided by the Alfred P. Sloan Foundation, the Participating Institutions, the National Science Foundation, the U.S. Department of Energy, the National Aeronautics and Space Administration, the Japanese Monbukagakusho, the Max Planck Society, and the Higher Education Funding Council for England.
A list of participating institutions can be obtained from the SDSS
Web Site http://www.sdss.org/.

\section*{Data Availability}

The data that support the findings of this study are available on request from the corresponding author, A.L.B.R.,  upon reasonable request.



\bibliographystyle{mnras}
\bibliography{refs}

\begin{thebibliography}{}
\makeatletter
\relax
\def\mn@urlcharsother{\let\do\@makeother \do\$\do\&\do\#\do\^\do\_\do\%\do\~}
\def\mn@doi{\begingroup\mn@urlcharsother \@ifnextchar [ {\mn@doi@}
  {\mn@doi@[]}}
\def\mn@doi@[#1]#2{\def\@tempa{#1}\ifx\@tempa\@empty \href
  {http://dx.doi.org/#2} {doi:#2}\else \href {http://dx.doi.org/#2} {#1}\fi
  \endgroup}
\def\mn@eprint#1#2{\mn@eprint@#1:#2::\@nil}
\def\mn@eprint@arXiv#1{\href {http://arxiv.org/abs/#1} {{\tt arXiv:#1}}}
\def\mn@eprint@dblp#1{\href {http://dblp.uni-trier.de/rec/bibtex/#1.xml}
  {dblp:#1}}
\def\mn@eprint@#1:#2:#3:#4\@nil{\def\@tempa {#1}\def\@tempb {#2}\def\@tempc
  {#3}\ifx \@tempc \@empty \let \@tempc \@tempb \let \@tempb \@tempa \fi \ifx
  \@tempb \@empty \def\@tempb {arXiv}\fi \@ifundefined
  {mn@eprint@\@tempb}{\@tempb:\@tempc}{\expandafter \expandafter \csname
  mn@eprint@\@tempb\endcsname \expandafter{\@tempc}}}

\bibitem[\protect\citeauthoryear{{Adami}, {Mazure}, {Biviano}, {Katgert}  \&
  {Rhee}}{{Adami} et~al.}{1998a}]{A98}
{Adami} C.,  {Mazure} A.,  {Biviano} A.,  {Katgert} P.,   {Rhee} G.,  1998a,
  \aap, \href {https://ui.adsabs.harvard.edu/abs/1998A&A...331..493A} {331,
  493}

\bibitem[\protect\citeauthoryear{Adami, Mazure, Katgert  \& Biviano}{Adami
  et~al.}{1998b}]{adami1998eso}
Adami C.,  Mazure A.,  Katgert P.,   Biviano A.,  1998b, Astronomy and
  Astrophysics, 336, 63

\bibitem[\protect\citeauthoryear{Aguado et~al.,}{Aguado
  et~al.}{2019}]{aguado2019fifteenth}
Aguado D.~S.,  et~al., 2019, The Astrophysical Journal Supplement Series, 240,
  23

\bibitem[\protect\citeauthoryear{{Aguerri}, {S{\'a}nchez-Janssen}  \&
  {Mu{\~n}oz-Tu{\~n}{\'o}n}}{{Aguerri} et~al.}{2007}]{aguerri07}
{Aguerri} J.~A.~L.,  {S{\'a}nchez-Janssen} R.,   {Mu{\~n}oz-Tu{\~n}{\'o}n} C.,
  2007, \mn@doi [\aap] {10.1051/0004-6361:20066478}, \href
  {https://ui.adsabs.harvard.edu/abs/2007A&A...471...17A} {471, 17}

\bibitem[\protect\citeauthoryear{Alabi et~al.,}{Alabi
  et~al.}{2017}]{alabi2017sluggs}
Alabi A.~B.,  et~al., 2017, Monthly Notices of the Royal Astronomical Society,
  468, 3949

\bibitem[\protect\citeauthoryear{Allen et~al.,}{Allen
  et~al.}{2015}]{allen2015differential}
Allen R.~J.,  et~al., 2015, The Astrophysical Journal, 806, 3

\bibitem[\protect\citeauthoryear{Andreon}{Andreon}{2018}]{andreon2018cosmic}
Andreon S.,  2018, Astronomy \& Astrophysics, 617, A53

\bibitem[\protect\citeauthoryear{Ascaso, Aguerri, Varela, Cava, Bettoni, Moles
  \& D'Onofrio}{Ascaso et~al.}{2010}]{ascaso2010evolution}
Ascaso B.,  Aguerri J.,  Varela J.,  Cava A.,  Bettoni D.,  Moles M.,
  D'Onofrio M.,  2010, The Astrophysical Journal, 726, 69

\bibitem[\protect\citeauthoryear{Barchi et~al.,}{Barchi
  et~al.}{2020}]{barchi2020machine}
Barchi P.,  et~al., 2020, Astronomy and Computing, 30, 100334

\bibitem[\protect\citeauthoryear{{Berlind} et~al.,}{{Berlind}
  et~al.}{2006}]{B06}
{Berlind} A.~A.,  et~al., 2006, \mn@doi [\apjs] {10.1086/508170}, \href
  {https://ui.adsabs.harvard.edu/abs/2006ApJS..167....1B} {167, 1}

\bibitem[\protect\citeauthoryear{Bernardi}{Bernardi}{2009}]{bernardi2009evolution}
Bernardi M.,  2009, Monthly Notices of the Royal Astronomical Society, 395,
  1491

\bibitem[\protect\citeauthoryear{Bernardi et~al.,}{Bernardi
  et~al.}{2003}]{bernardi2003early}
Bernardi M.,  et~al., 2003, The Astronomical Journal, 125, 1849

\bibitem[\protect\citeauthoryear{Berrier, Stewart, Bullock, Purcell, Barton  \&
  Wechsler}{Berrier et~al.}{2008}]{berrier2008assembly}
Berrier J.~C.,  Stewart K.~R.,  Bullock J.~S.,  Purcell C.~W.,  Barton E.~J.,
  Wechsler R.~H.,  2008, The Astrophysical Journal, 690, 1292

\bibitem[\protect\citeauthoryear{Binney \& Tremaine}{Binney \&
  Tremaine}{1987}]{binney1987galactic}
Binney J.,  Tremaine S.,  1987, Princeton, NJ, Princeton University Press,
  1987, 747

\bibitem[\protect\citeauthoryear{{Biviano}, {Katgert}, {Thomas}  \&
  {Adami}}{{Biviano} et~al.}{2002}]{biviano2002eso}
{Biviano} A.,  {Katgert} P.,  {Thomas} T.,   {Adami} C.,  2002, \mn@doi [\aap]
  {10.1051/0004-6361:20020340}, \href
  {https://ui.adsabs.harvard.edu/abs/2002A&A...387....8B} {387, 8}

\bibitem[\protect\citeauthoryear{{Biviano}, {Murante}, {Borgani}, {Diaferio},
  {Dolag}  \& {Girardi}}{{Biviano} et~al.}{2006}]{Bi06}
{Biviano} A.,  {Murante} G.,  {Borgani} S.,  {Diaferio} A.,  {Dolag} K.,
  {Girardi} M.,  2006, \mn@doi [\aap] {10.1051/0004-6361:20064918}, \href
  {https://ui.adsabs.harvard.edu/abs/2006A&A...456...23B} {456, 23}

\bibitem[\protect\citeauthoryear{Blanton \& Moustakas}{Blanton \&
  Moustakas}{2009}]{blanton2009physical}
Blanton M.~R.,  Moustakas J.,  2009, Annual Review of Astronomy and
  Astrophysics, 47, 159

\bibitem[\protect\citeauthoryear{Boselli \& Gavazzi}{Boselli \&
  Gavazzi}{2006}]{boselli2006environmental}
Boselli A.,  Gavazzi G.,  2006, Publications of the Astronomical Society of the
  Pacific, 118, 517

\bibitem[\protect\citeauthoryear{Buitrago, Trujillo, Curtis-Lake, Montes,
  Cooper, Bruce, P{\'e}rez-Gonz{\'a}lez  \& Cirasuolo}{Buitrago
  et~al.}{2017}]{buitrago2017cosmic}
Buitrago F.,  Trujillo I.,  Curtis-Lake E.,  Montes M.,  Cooper A.~P.,  Bruce
  V.~A.,  P{\'e}rez-Gonz{\'a}lez P.~G.,   Cirasuolo M.,  2017, Monthly Notices
  of the Royal Astronomical Society, 466, 4888

\bibitem[\protect\citeauthoryear{Cappellari}{Cappellari}{2016}]{cappellari2016structure}
Cappellari M.,  2016, Annual review of astronomy and astrophysics, 54, 597

\bibitem[\protect\citeauthoryear{Cappellari et~al.,}{Cappellari
  et~al.}{2006}]{cappellari2006sauron}
Cappellari M.,  et~al., 2006, Monthly Notices of the Royal Astronomical
  Society, 366, 1126

\bibitem[\protect\citeauthoryear{{Cava} et~al.,}{{Cava} et~al.}{2017}]{cava17}
{Cava} A.,  et~al., 2017, \mn@doi [\aap] {10.1051/0004-6361/201730785}, \href
  {https://ui.adsabs.harvard.edu/abs/2017A&A...606A.108C} {606, A108}

\bibitem[\protect\citeauthoryear{Chandrasekhar}{Chandrasekhar}{1943}]{chandrasekhar1943dynamical}
Chandrasekhar S.,  1943, Astrophysical Journal, 97, 255

\bibitem[\protect\citeauthoryear{Cimatti, Fraternali  \& Nipoti}{Cimatti
  et~al.}{2019}]{cimatti2019introduction}
Cimatti A.,  Fraternali F.,   Nipoti C.,  2019, Introduction to Galaxy
  Formation and Evolution: From Primordial Gas to Present-Day Galaxies.
Cambridge University Press

\bibitem[\protect\citeauthoryear{Coenda, Rios, Muriel, Cora, Mart{\'\i}nez,
  Ruiz  \& Vega-Mart{\'\i}nez}{Coenda et~al.}{2022}]{coenda2022reconstructing}
Coenda V.,  Rios M. d.~l.,  Muriel H.,  Cora S.~A.,  Mart{\'\i}nez H.~J.,  Ruiz
  A.~N.,   Vega-Mart{\'\i}nez C.~A.,  2022, Monthly Notices of the Royal
  Astronomical Society, 510, 1934

\bibitem[\protect\citeauthoryear{Conover \& Iman}{Conover \&
  Iman}{1979}]{conover1979multiple}
Conover W.~J.,  Iman R.~L.,  1979, Technical report, Multiple-comparisons
  procedures. Informal report.
Los Alamos National Lab.(LANL), Los Alamos, NM (United States)

\bibitem[\protect\citeauthoryear{Conroy, van Dokkum  \& Kravtsov}{Conroy
  et~al.}{2015}]{conroy2015preventing}
Conroy C.,  van Dokkum P.~G.,   Kravtsov A.,  2015, The Astrophysical Journal,
  803, 77

\bibitem[\protect\citeauthoryear{Cypriano, Sodr{\'e}~Jr, Campusano, Dale  \&
  Hardy}{Cypriano et~al.}{2006}]{cypriano2006shrinking}
Cypriano E.~S.,  Sodr{\'e}~Jr L.,  Campusano L.~E.,  Dale D.~A.,   Hardy E.,
  2006, The Astronomical Journal, 131, 2417

\bibitem[\protect\citeauthoryear{Daddi et~al.,}{Daddi
  et~al.}{2005}]{daddi2005passively}
Daddi E.,  et~al., 2005, The Astrophysical Journal, 626, 680

\bibitem[\protect\citeauthoryear{Damjanov et~al.,}{Damjanov
  et~al.}{2009}]{damjanov2009red}
Damjanov I.,  et~al., 2009, The Astrophysical Journal, 695, 101

\bibitem[\protect\citeauthoryear{Damjanov et~al.,}{Damjanov
  et~al.}{2011}]{damjanov2011red}
Damjanov I.,  et~al., 2011, The Astrophysical Journal Letters, 739, L44

\bibitem[\protect\citeauthoryear{Damjanov, Zahid, Geller  \& Hwang}{Damjanov
  et~al.}{2015}]{damjanov2015environment}
Damjanov I.,  Zahid H.~J.,  Geller M.~J.,   Hwang H.~S.,  2015, The
  Astrophysical Journal, 815, 104

\bibitem[\protect\citeauthoryear{De~Propris, Bremer  \& Phillipps}{De~Propris
  et~al.}{2016}]{de2016morphological}
De~Propris R.,  Bremer M.~N.,   Phillipps S.,  2016, Monthly Notices of the
  Royal Astronomical Society, 461, 4517

\bibitem[\protect\citeauthoryear{Di~Cintio, Tremmel, Governato, Pontzen,
  Zavala, Bastidas~Fry, Brooks  \& Vogelsberger}{Di~Cintio
  et~al.}{2017}]{di2017rumble}
Di~Cintio A.,  Tremmel M.,  Governato F.,  Pontzen A.,  Zavala J.,
  Bastidas~Fry A.,  Brooks A.,   Vogelsberger M.,  2017, Monthly Notices of the
  Royal Astronomical Society, 469, 2845

\bibitem[\protect\citeauthoryear{Di~Matteo, Pipino, Lehnert, Combes  \&
  Semelin}{Di~Matteo et~al.}{2009}]{di2009survival}
Di~Matteo P.,  Pipino A.,  Lehnert M.~D.,  Combes F.,   Semelin B.,  2009,
  Astronomy \& Astrophysics, 499, 427

\bibitem[\protect\citeauthoryear{{Dom{\'\i}nguez S{\'a}nchez},
  {Huertas-Company}, {Bernardi}, {Tuccillo}  \& {Fischer}}{{Dom{\'\i}nguez
  S{\'a}nchez} et~al.}{2018}]{dominguez2018improving}
{Dom{\'\i}nguez S{\'a}nchez} H.,  {Huertas-Company} M.,  {Bernardi} M.,
  {Tuccillo} D.,   {Fischer} J.~L.,  2018, \mn@doi [\mnras]
  {10.1093/mnras/sty338}, \href
  {https://ui.adsabs.harvard.edu/abs/2018MNRAS.476.3661D} {476, 3661}

\bibitem[\protect\citeauthoryear{Dressler \& Shectman}{Dressler \&
  Shectman}{1988}]{dressler1988evidence}
Dressler A.,  Shectman S.~A.,  1988, The Astronomical Journal, 95, 985

\bibitem[\protect\citeauthoryear{Edwards \& Patton}{Edwards \&
  Patton}{2012}]{edwards2012close}
Edwards L.~O.,  Patton D.~R.,  2012, Monthly Notices of the Royal Astronomical
  Society, 425, 287

\bibitem[\protect\citeauthoryear{Evrard et~al.,}{Evrard
  et~al.}{2008}]{evrard2008virial}
Evrard A.~E.,  et~al., 2008, The astrophysical journal, 672, 122

\bibitem[\protect\citeauthoryear{Faber \& Jackson}{Faber \&
  Jackson}{1976}]{faber1976velocity}
Faber S.~M.,  Jackson R.~E.,  1976, The Astrophysical Journal, 204, 668

\bibitem[\protect\citeauthoryear{Fadda, Girardi, Giuricin, Mardirossian  \&
  Mezzetti}{Fadda et~al.}{1996}]{fadda1996observational}
Fadda D.,  Girardi M.,  Giuricin G.,  Mardirossian F.,   Mezzetti M.,  1996,
  The Astrophysical Journal, 473, 670

\bibitem[\protect\citeauthoryear{Feldmann, Carollo  \& Mayer}{Feldmann
  et~al.}{2011}]{feldmann2011hubble}
Feldmann R.,  Carollo C.~M.,   Mayer L.,  2011, The Astrophysical Journal, 736,
  88

\bibitem[\protect\citeauthoryear{Gargiulo, Saracco, Longhetti, La~Barbera  \&
  Tamburri}{Gargiulo et~al.}{2012}]{gargiulo2012spatially}
Gargiulo A.,  Saracco P.,  Longhetti M.,  La~Barbera F.,   Tamburri S.,  2012,
  Monthly Notices of the Royal Astronomical Society, 425, 2698

\bibitem[\protect\citeauthoryear{Gill, Knebe  \& Gibson}{Gill
  et~al.}{2005}]{gill2005evolution}
Gill S.~P.,  Knebe A.,   Gibson B.~K.,  2005, Monthly Notices of the Royal
  Astronomical Society, 356, 1327

\bibitem[\protect\citeauthoryear{{Girardi}, {Giuricin}, {Mardirossian},
  {Mezzetti}  \& {Boschin}}{{Girardi} et~al.}{1998}]{G98}
{Girardi} M.,  {Giuricin} G.,  {Mardirossian} F.,  {Mezzetti} M.,   {Boschin}
  W.,  1998, \mn@doi [\apj] {10.1086/306157}, \href
  {https://ui.adsabs.harvard.edu/abs/1998ApJ...505...74G} {505, 74}

\bibitem[\protect\citeauthoryear{Goto}{Goto}{2005}]{goto2005velocity}
Goto T.,  2005, Monthly Notices of the Royal Astronomical Society, 359, 1415

\bibitem[\protect\citeauthoryear{Graves, Faber  \& Schiavon}{Graves
  et~al.}{2009}]{graves2009dissecting}
Graves G.~J.,  Faber S.,   Schiavon R.~P.,  2009, The Astrophysical Journal,
  693, 486

\bibitem[\protect\citeauthoryear{Guo, White, Li  \& Boylan-Kolchin}{Guo
  et~al.}{2010}]{guo2010galaxies}
Guo Q.,  White S.,  Li C.,   Boylan-Kolchin M.,  2010, Monthly Notices of the
  Royal Astronomical Society, 404, 1111

\bibitem[\protect\citeauthoryear{Hilz, Naab  \& Ostriker}{Hilz
  et~al.}{2013}]{hilz2013minor}
Hilz M.,  Naab T.,   Ostriker J.~P.,  2013, Monthly Notices of the Royal
  Astronomical Society, 429, 2924

\bibitem[\protect\citeauthoryear{Huang, Chen, Johnson  \& Weiner}{Huang
  et~al.}{2016}]{huang2016characterizing}
Huang Y.-H.,  Chen H.-W.,  Johnson S.~D.,   Weiner B.~J.,  2016, Monthly
  Notices of the Royal Astronomical Society, 455, 1713

\bibitem[\protect\citeauthoryear{Hyde \& Bernardi}{Hyde \&
  Bernardi}{2009}]{hyde2009luminosity}
Hyde J.~B.,  Bernardi M.,  2009, Monthly Notices of the Royal Astronomical
  Society, 396, 1171

\bibitem[\protect\citeauthoryear{Iodice et~al.,}{Iodice
  et~al.}{2017}]{iodice2017fornax}
Iodice E.,  et~al., 2017, The Astrophysical Journal, 839, 21

\bibitem[\protect\citeauthoryear{Johansson, Naab  \& Ostriker}{Johansson
  et~al.}{2012}]{johansson2012forming}
Johansson P.~H.,  Naab T.,   Ostriker J.~P.,  2012, The Astrophysical Journal,
  754, 115

\bibitem[\protect\citeauthoryear{Joshi, Wadsley  \& Parker}{Joshi
  et~al.}{2017}]{joshi2017preprocessing}
Joshi G.~D.,  Wadsley J.,   Parker L.~C.,  2017, Monthly Notices of the Royal
  Astronomical Society, 468, 4625

\bibitem[\protect\citeauthoryear{{Kauffmann}, {White}, {Heckman}, {M{\'e}nard},
  {Brinchmann}, {Charlot}, {Tremonti}  \& {Brinkmann}}{{Kauffmann}
  et~al.}{2004a}]{2004MNRAS.353..713K}
{Kauffmann} G.,  {White} S.~D.~M.,  {Heckman} T.~M.,  {M{\'e}nard} B.,
  {Brinchmann} J.,  {Charlot} S.,  {Tremonti} C.,   {Brinkmann} J.,  2004a,
  \mn@doi [\mnras] {10.1111/j.1365-2966.2004.08117.x}, \href
  {http://adsabs.harvard.edu/abs/2004MNRAS.353..713K} {353, 713}

\bibitem[\protect\citeauthoryear{Kauffmann, White, Heckman, M{\'e}nard,
  Brinchmann, Charlot, Tremonti  \& Brinkmann}{Kauffmann
  et~al.}{2004b}]{kauffmann2004environmental}
Kauffmann G.,  White S.~D.,  Heckman T.~M.,  M{\'e}nard B.,  Brinchmann J.,
  Charlot S.,  Tremonti C.,   Brinkmann J.,  2004b, Monthly Notices of the
  Royal Astronomical Society, 353, 713

\bibitem[\protect\citeauthoryear{Kuchner et~al.,}{Kuchner
  et~al.}{2022}]{kuchner2022inventory}
Kuchner U.,  et~al., 2022, Monthly Notices of the Royal Astronomical Society,
  510, 581

\bibitem[\protect\citeauthoryear{La~Barbera \& De~Carvalho}{La~Barbera \&
  De~Carvalho}{2009}]{la2009origin}
La~Barbera F.,  De~Carvalho R.,  2009, The Astrophysical Journal Letters, 699,
  L76

\bibitem[\protect\citeauthoryear{{La Barbera}, {Lopes}, {de Carvalho}, {de La
  Rosa}  \& {Berlind}}{{La Barbera} et~al.}{2010}]{Lab10}
{La Barbera} F.,  {Lopes} P.~A.~A.,  {de Carvalho} R.~R.,  {de La Rosa} I.~G.,
   {Berlind} A.~A.,  2010, \mn@doi [\mnras] {10.1111/j.1365-2966.2010.17273.x},
  \href {https://ui.adsabs.harvard.edu/abs/2010MNRAS.408.1361L} {408, 1361}

\bibitem[\protect\citeauthoryear{Lani et~al.,}{Lani
  et~al.}{2013}]{lani2013evidence}
Lani C.,  et~al., 2013, Monthly Notices of the Royal Astronomical Society, 435,
  207

\bibitem[\protect\citeauthoryear{{Lintott} et~al.,}{{Lintott}
  et~al.}{2008}]{2008MNRAS.389.1179L}
{Lintott} C.~J.,  et~al., 2008, \mn@doi [\mnras]
  {10.1111/j.1365-2966.2008.13689.x}, \href
  {http://adsabs.harvard.edu/abs/2008MNRAS.389.1179L} {389, 1179}

\bibitem[\protect\citeauthoryear{Liu, Mao, Deng, Xia  \& Wen}{Liu
  et~al.}{2009}]{liu2009major}
Liu F.,  Mao S.,  Deng Z.,  Xia X.,   Wen Z.,  2009, Monthly Notices of the
  Royal Astronomical Society, 396, 2003

\bibitem[\protect\citeauthoryear{Liu, Lei, Meng  \& Jiang}{Liu
  et~al.}{2015}]{liu2015ongoing}
Liu F.,  Lei F.,  Meng X.,   Jiang D.,  2015, Monthly Notices of the Royal
  Astronomical Society, 447, 1491

\bibitem[\protect\citeauthoryear{{\L}okas}{{\L}okas}{2020}]{lokas2020tidal}
{\L}okas E.~L.,  2020, Astronomy \& Astrophysics, 638, A133

\bibitem[\protect\citeauthoryear{{Lopes}}{{Lopes}}{2007}]{L07}
{Lopes} P.~A.~A.,  2007, \mn@doi [\mnras] {10.1111/j.1365-2966.2007.12203.x},
  \href {https://ui.adsabs.harvard.edu/abs/2007MNRAS.380.1608L} {380, 1608}

\bibitem[\protect\citeauthoryear{{Lopes}, {de Carvalho}, {Kohl-Moreira}  \&
  {Jones}}{{Lopes} et~al.}{2009}]{L09}
{Lopes} P.~A.~A.,  {de Carvalho} R.~R.,  {Kohl-Moreira} J.~L.,   {Jones} C.,
  2009, \mn@doi [\mnras] {10.1111/j.1365-2966.2008.13962.x}, \href
  {https://ui.adsabs.harvard.edu/abs/2009MNRAS.392..135L} {392, 135}

\bibitem[\protect\citeauthoryear{L{\'o}pez-Sanjuan et~al.,}{L{\'o}pez-Sanjuan
  et~al.}{2012}]{lopez2012dominant}
L{\'o}pez-Sanjuan C.,  et~al., 2012, Astronomy \& Astrophysics, 548, A7

\bibitem[\protect\citeauthoryear{Mamon, Cava, Biviano, Moretti, Poggianti  \&
  Bettoni}{Mamon et~al.}{2019}]{mamon2019structural}
Mamon G.,  Cava A.,  Biviano A.,  Moretti A.,  Poggianti B.,   Bettoni D.,
  2019, \aap, 631, A131

\bibitem[\protect\citeauthoryear{Mancillas, Duc, Combes, Bournaud, Emsellem,
  Martig  \& Michel-Dansac}{Mancillas et~al.}{2019}]{mancillas2019probing}
Mancillas B.,  Duc P.-A.,  Combes F.,  Bournaud F.,  Emsellem E.,  Martig M.,
  Michel-Dansac L.,  2019, Astronomy \& Astrophysics, 632, A122

\bibitem[\protect\citeauthoryear{Matharu et~al.,}{Matharu
  et~al.}{2019}]{matharu2019hst}
Matharu J.,  et~al., 2019, Monthly Notices of the Royal Astronomical Society,
  484, 595

\bibitem[\protect\citeauthoryear{Matteuzzi, Marinacci, Nipoti  \&
  Andreon}{Matteuzzi et~al.}{2022}]{matteuzzi2022newcomers}
Matteuzzi M.,  Marinacci F.,  Nipoti C.,   Andreon S.,  2022, Monthly Notices
  of the Royal Astronomical Society, 513, 3893

\bibitem[\protect\citeauthoryear{McIntosh, Guo, Hertzberg, Katz, Mo, Van
  Den~Bosch  \& Yang}{McIntosh et~al.}{2008}]{mcintosh2008ongoing}
McIntosh D.~H.,  Guo Y.,  Hertzberg J.,  Katz N.,  Mo H.,  Van Den~Bosch F.~C.,
    Yang X.,  2008, Monthly Notices of the Royal Astronomical Society, 388,
  1537

\bibitem[\protect\citeauthoryear{Miller, van~den Bosch, Green  \& Ogiya}{Miller
  et~al.}{2020}]{miller2020dynamical}
Miller T.~B.,  van~den Bosch F.~C.,  Green S.~B.,   Ogiya G.,  2020, Monthly
  Notices of the Royal Astronomical Society, 495, 4496

\bibitem[\protect\citeauthoryear{Morell, Ribeiro, de Carvalho, Rembold, Lopes
  \& Costa}{Morell et~al.}{2020}]{morell2020classification}
Morell D.,  Ribeiro A.,  de Carvalho R.,  Rembold S.,  Lopes P.,   Costa A.,
  2020, Monthly Notices of the Royal Astronomical Society, 494, 3317

\bibitem[\protect\citeauthoryear{Muriel \& Coenda}{Muriel \&
  Coenda}{2014}]{muriel2014galaxy}
Muriel H.,  Coenda V.,  2014, Astronomy \& Astrophysics, 564, A85

\bibitem[\protect\citeauthoryear{Naab, Johansson  \& Ostriker}{Naab
  et~al.}{2009}]{naab2009minor}
Naab T.,  Johansson P.~H.,   Ostriker J.~P.,  2009, The Astrophysical Journal
  Letters, 699, L178

\bibitem[\protect\citeauthoryear{{Nair} \& {Abraham}}{{Nair} \&
  {Abraham}}{2010}]{nair2010catalog}
{Nair} P.~B.,  {Abraham} R.~G.,  2010, \mn@doi [\apjs]
  {10.1088/0067-0049/186/2/427}, \href
  {https://ui.adsabs.harvard.edu/abs/2010ApJS..186..427N} {186, 427}

\bibitem[\protect\citeauthoryear{Nascimento, Lopes, Ribeiro, Costa  \&
  Morell}{Nascimento et~al.}{2019}]{nascimento2019influence}
Nascimento R.~S.,  Lopes P.~A.,  Ribeiro A.~L.,  Costa A.~P.,   Morell D.~F.,
  2019, Monthly Notices of the Royal Astronomical Society: Letters, 483, L121

\bibitem[\protect\citeauthoryear{Newman, Ellis, Treu  \& Bundy}{Newman
  et~al.}{2010}]{newman2010keck}
Newman A.~B.,  Ellis R.~S.,  Treu T.,   Bundy K.,  2010, The Astrophysical
  Journal Letters, 717, L103

\bibitem[\protect\citeauthoryear{Nigoche-Netro, Ramos-Larios, Lagos, De~la
  Fuente, Ruelas-Mayorga, Mendez-Abreu, Kemp  \& Diaz}{Nigoche-Netro
  et~al.}{2019}]{nigoche2019quantity}
Nigoche-Netro A.,  Ramos-Larios G.,  Lagos P.,  De~la Fuente E.,
  Ruelas-Mayorga A.,  Mendez-Abreu J.,  Kemp S.,   Diaz R.,  2019, Monthly
  Notices of the Royal Astronomical Society, 488, 1320

\bibitem[\protect\citeauthoryear{Nipoti}{Nipoti}{2017}]{nipoti2017special}
Nipoti C.,  2017, Monthly Notices of the Royal Astronomical Society, 467, 661

\bibitem[\protect\citeauthoryear{Old et~al.,}{Old et~al.}{2015}]{old2015galaxy}
Old L.,  et~al., 2015, Monthly Notices of the Royal Astronomical Society, 449,
  1897

\bibitem[\protect\citeauthoryear{Oogi, Habe  \& Ishiyama}{Oogi
  et~al.}{2016}]{oogi2016mass}
Oogi T.,  Habe A.,   Ishiyama T.,  2016, Monthly Notices of the Royal
  Astronomical Society, 456, 300

\bibitem[\protect\citeauthoryear{Oser, Ostriker, Naab, Johansson  \&
  Burkert}{Oser et~al.}{2010}]{oser2010two}
Oser L.,  Ostriker J.~P.,  Naab T.,  Johansson P.~H.,   Burkert A.,  2010, The
  Astrophysical Journal, 725, 2312

\bibitem[\protect\citeauthoryear{Ownsworth, Conselice, Mortlock, Hartley,
  Almaini, Duncan  \& Mundy}{Ownsworth et~al.}{2014}]{ownsworth2014minor}
Ownsworth J.~R.,  Conselice C.~J.,  Mortlock A.,  Hartley W.~G.,  Almaini O.,
  Duncan K.,   Mundy C.~J.,  2014, Monthly Notices of the Royal Astronomical
  Society, 445, 2198

\bibitem[\protect\citeauthoryear{{Park} \& {Choi}}{{Park} \&
  {Choi}}{2005}]{parkchoi}
{Park} C.,  {Choi} Y.-Y.,  2005, \mn@doi [\apjl] {10.1086/499243}, \href
  {http://adsabs.harvard.edu/abs/2005ApJ...635L..29P} {635, L29}

\bibitem[\protect\citeauthoryear{{Pasquali}, {Smith}, {Gallazzi}, {De Lucia},
  {Zibetti}, {Hirschmann}  \& {Yi}}{{Pasquali} et~al.}{2019}]{pasquali19}
{Pasquali} A.,  {Smith} R.,  {Gallazzi} A.,  {De Lucia} G.,  {Zibetti} S.,
  {Hirschmann} M.,   {Yi} S.~K.,  2019, \mn@doi [\mnras]
  {10.1093/mnras/sty3530}, \href
  {https://ui.adsabs.harvard.edu/abs/2019MNRAS.484.1702P} {484, 1702}

\bibitem[\protect\citeauthoryear{Peletier, Davies, Illingworth, Davis  \&
  Cawson}{Peletier et~al.}{1990}]{peletier1990ccd}
Peletier R.~F.,  Davies R.~L.,  Illingworth G.~D.,  Davis L.~E.,   Cawson M.,
  1990, The Astronomical Journal, 100, 1091

\bibitem[\protect\citeauthoryear{Peng et~al.,}{Peng
  et~al.}{2010}]{peng2010mass}
Peng Y.-j.,  et~al., 2010, The Astrophysical Journal, 721, 193

\bibitem[\protect\citeauthoryear{Pillepich et~al.,}{Pillepich
  et~al.}{2018}]{pillepich2018first}
Pillepich A.,  et~al., 2018, Monthly Notices of the Royal Astronomical Society,
  475, 648

\bibitem[\protect\citeauthoryear{{Popesso}, {Biviano}, {B{\"o}hringer},
  {Romaniello}  \& {Voges}}{{Popesso} et~al.}{2005}]{P05}
{Popesso} P.,  {Biviano} A.,  {B{\"o}hringer} H.,  {Romaniello} M.,   {Voges}
  W.,  2005, \mn@doi [\aap] {10.1051/0004-6361:20041915}, \href
  {https://ui.adsabs.harvard.edu/abs/2005A&A...433..431P} {433, 431}

\bibitem[\protect\citeauthoryear{{Popesso}, {Biviano}, {B{\"o}hringer}  \&
  {Romaniello}}{{Popesso} et~al.}{2007}]{P07}
{Popesso} P.,  {Biviano} A.,  {B{\"o}hringer} H.,   {Romaniello} M.,  2007,
  \mn@doi [\aap] {10.1051/0004-6361:20054708}, \href
  {https://ui.adsabs.harvard.edu/abs/2007A&A...464..451P} {464, 451}

\bibitem[\protect\citeauthoryear{Poveda}{Poveda}{1958}]{poveda1958masses}
Poveda A.,  1958, Boletin de los Observatorios Tonantzintla y Tacubaya, 2

\bibitem[\protect\citeauthoryear{Rasmussen, Mulchaey, Bai, Ponman, Raychaudhury
   \& Dariush}{Rasmussen et~al.}{2010}]{rasmussen2010witnessing}
Rasmussen J.,  Mulchaey J.~S.,  Bai L.,  Ponman T.~J.,  Raychaudhury S.,
  Dariush A.,  2010, The Astrophysical Journal, 717, 958

\bibitem[\protect\citeauthoryear{Renzini}{Renzini}{2006}]{renzini2006stellar}
Renzini A.,  2006, Annu. Rev. Astron. Astrophys., 44, 141

\bibitem[\protect\citeauthoryear{{Rhee}, {Smith}, {Choi}, {Yi}, {Jaff{\'e}},
  {Candlish}  \& {S{\'a}nchez-J{\'a}nssen}}{{Rhee}
  et~al.}{2017}]{rhee2017phase}
{Rhee} J.,  {Smith} R.,  {Choi} H.,  {Yi} S.~K.,  {Jaff{\'e}} Y.,  {Candlish}
  G.,   {S{\'a}nchez-J{\'a}nssen} R.,  2017, \mn@doi [\apj]
  {10.3847/1538-4357/aa6d6c}, \href
  {https://ui.adsabs.harvard.edu/abs/2017ApJ...843..128R} {843, 128}

\bibitem[\protect\citeauthoryear{Rhee, Smith, Choi, Contini, Jung, Han  \&
  Sukyoung}{Rhee et~al.}{2020}]{rhee2020yzics}
Rhee J.,  Smith R.,  Choi H.,  Contini E.,  Jung S.~L.,  Han S.,   Sukyoung
  K.~Y.,  2020, The Astrophysical Journal Supplement Series, 247, 45

\bibitem[\protect\citeauthoryear{{Ribeiro}, {de Carvalho}, {Trevisan},
  {Capelato}, {La Barbera}, {Lopes}  \& {Schilling}}{{Ribeiro}
  et~al.}{2013}]{2013MNRAS.434..784R}
{Ribeiro} A.~L.~B.,  {de Carvalho} R.~R.,  {Trevisan} M.,  {Capelato} H.~V.,
  {La Barbera} F.,  {Lopes} P.~A.~A.,   {Schilling} A.~C.,  2013, \mn@doi
  [\mnras] {10.1093/mnras/stt1071}, \href
  {http://adsabs.harvard.edu/abs/2013MNRAS.434..784R} {434, 784}

\bibitem[\protect\citeauthoryear{Rocha, Peter, Bullock, Kaplinghat,
  Garrison-Kimmel, Onorbe  \& Moustakas}{Rocha
  et~al.}{2013}]{rocha2013cosmological}
Rocha M.,  Peter A.~H.,  Bullock J.~S.,  Kaplinghat M.,  Garrison-Kimmel S.,
  Onorbe J.,   Moustakas L.~A.,  2013, Monthly Notices of the Royal
  Astronomical Society, 430, 81

\bibitem[\protect\citeauthoryear{Rodriguez-Gomez et~al.,}{Rodriguez-Gomez
  et~al.}{2016}]{rodriguez2016stellar}
Rodriguez-Gomez V.,  et~al., 2016, Monthly Notices of the Royal Astronomical
  Society, 458, 2371

\bibitem[\protect\citeauthoryear{Roy et~al.,}{Roy
  et~al.}{2018}]{roy2018evolution}
Roy N.,  et~al., 2018, Monthly Notices of the Royal Astronomical Society, 480,
  1057

\bibitem[\protect\citeauthoryear{Sampaio, de Carvalho, Ferreras, Lagan{\'a},
  Ribeiro  \& Rembold}{Sampaio et~al.}{2021}]{sampaio2021investigating}
Sampaio V.,  de Carvalho R.,  Ferreras I.,  Lagan{\'a} T.,  Ribeiro A.,
  Rembold S.,  2021, Monthly Notices of the Royal Astronomical Society, 503,
  3065

\bibitem[\protect\citeauthoryear{Saracco, Gargiulo, Ciocca  \&
  Marchesini}{Saracco et~al.}{2017}]{saracco2017cluster}
Saracco P.,  Gargiulo A.,  Ciocca F.,   Marchesini D.,  2017, Astronomy \&
  Astrophysics, 597, A122

\bibitem[\protect\citeauthoryear{Schechter}{Schechter}{2016}]{schechter2016new}
Schechter P.~L.,  2016, The General Assembly of Galaxy Halos: Structure, Origin
  and Evolution, 317, 35

\bibitem[\protect\citeauthoryear{Shankar, Marulli, Bernardi, Mei, Meert  \&
  Vikram}{Shankar et~al.}{2013}]{shankar2013size}
Shankar F.,  Marulli F.,  Bernardi M.,  Mei S.,  Meert A.,   Vikram V.,  2013,
  Monthly Notices of the Royal Astronomical Society, 428, 109

\bibitem[\protect\citeauthoryear{Shen, Mo, White, Blanton, Kauffmann, Voges,
  Brinkmann  \& Csabai}{Shen et~al.}{2003}]{shen2003size}
Shen S.,  Mo H.,  White S.~D.,  Blanton M.~R.,  Kauffmann G.,  Voges W.,
  Brinkmann J.,   Csabai I.,  2003, Monthly Notices of the Royal Astronomical
  Society, 343, 978

\bibitem[\protect\citeauthoryear{Signorell, Aho, Alfons, Anderegg, Aragon
  et~al.}{Signorell et~al.}{2016}]{signorell2016desctools}
Signorell A.,  Aho K.,  Alfons A.,  Anderegg N.,  Aragon T.,   et~al., 2016, R
  Foundation for Statistical Computing, Vienna, Austria

\bibitem[\protect\citeauthoryear{Simard, Mendel, Patton, Ellison  \&
  McConnachie}{Simard et~al.}{2011}]{simard2011catalog}
Simard L.,  Mendel J.~T.,  Patton D.~R.,  Ellison S.~L.,   McConnachie A.~W.,
  2011, The Astrophysical Journal Supplement Series, 196, 11

\bibitem[\protect\citeauthoryear{Smith, Choi, Lee, Rhee, Sanchez-Janssen  \&
  Sukyoung}{Smith et~al.}{2016}]{smith2016preferential}
Smith R.,  Choi H.,  Lee J.,  Rhee J.,  Sanchez-Janssen R.,   Sukyoung K.~Y.,
  2016, The Astrophysical Journal, 833, 109

\bibitem[\protect\citeauthoryear{Sodre~Jr, Capelato, Steiner  \&
  Mazure}{Sodre~Jr et~al.}{1989}]{sodre1989kinematical}
Sodre~Jr L.,  Capelato H.~V.,  Steiner J.~E.,   Mazure A.,  1989, The
  Astronomical Journal, 97, 1279

\bibitem[\protect\citeauthoryear{Song, Hwang, Park, Smith  \& Einasto}{Song
  et~al.}{2018}]{song2018redshift}
Song H.,  Hwang H.~S.,  Park C.,  Smith R.,   Einasto M.,  2018, The
  Astrophysical Journal, 869, 124

\bibitem[\protect\citeauthoryear{Stein, Jerjen  \& Federspiel}{Stein
  et~al.}{1997}]{stein1997velocity}
Stein P.,  Jerjen H.,   Federspiel M.,  1997, Astronomy and Astrophysics, 327,
  952

\bibitem[\protect\citeauthoryear{Tamfal, Mayer, Quinn, Capelo, Kazantzidis,
  Babul  \& Potter}{Tamfal et~al.}{2021}]{tamfal2021revisiting}
Tamfal T.,  Mayer L.,  Quinn T.~R.,  Capelo P.~R.,  Kazantzidis S.,  Babul A.,
   Potter D.,  2021, The Astrophysical Journal, 916, 55

\bibitem[\protect\citeauthoryear{Thomas, Maraston, Bender  \&
  De~Oliveira}{Thomas et~al.}{2005}]{thomas2005epochs}
Thomas D.,  Maraston C.,  Bender R.,   De~Oliveira C.~M.,  2005, The
  Astrophysical Journal, 621, 673

\bibitem[\protect\citeauthoryear{Tortora \& Napolitano}{Tortora \&
  Napolitano}{2022}]{tortora2022central}
Tortora C.,  Napolitano N.,  2022, Front. Astron. Space Sci. 8: 704419. doi:
  10.3389/fspas

\bibitem[\protect\citeauthoryear{Tortora, Napolitano, Roy, Radovich, Getman,
  Koopmans, Verdoes~Kleijn  \& Kuijken}{Tortora et~al.}{2018}]{tortora2018last}
Tortora C.,  Napolitano N.~R.,  Roy N.,  Radovich M.,  Getman F.,  Koopmans L.,
   Verdoes~Kleijn G.,   Kuijken K.,  2018, Monthly Notices of the Royal
  Astronomical Society, 473, 969

\bibitem[\protect\citeauthoryear{Tran, Moustakas, Gonzalez, Bai, Zaritsky  \&
  Kautsch}{Tran et~al.}{2008}]{tran2008late}
Tran K.-V.~H.,  Moustakas J.,  Gonzalez A.~H.,  Bai L.,  Zaritsky D.,   Kautsch
  S.~J.,  2008, The Astrophysical Journal, 683, L17

\bibitem[\protect\citeauthoryear{Trujillo, Conselice, Bundy, Cooper, Eisenhardt
   \& Ellis}{Trujillo et~al.}{2007}]{trujillo2007strong}
Trujillo I.,  Conselice C.~J.,  Bundy K.,  Cooper M.,  Eisenhardt P.,   Ellis
  R.~S.,  2007, Monthly Notices of the Royal Astronomical Society, 382, 109

\bibitem[\protect\citeauthoryear{Trujillo, Ferreras  \& de La~Rosa}{Trujillo
  et~al.}{2011}]{trujillo2011dissecting}
Trujillo I.,  Ferreras I.,   de La~Rosa I.~G.,  2011, Monthly Notices of the
  Royal Astronomical Society, 415, 3903

\bibitem[\protect\citeauthoryear{Van~Dokkum \& Conroy}{Van~Dokkum \&
  Conroy}{2010}]{van2010substantial}
Van~Dokkum P.~G.,  Conroy C.,  2010, Nature, 468, 940

\bibitem[\protect\citeauthoryear{Van~Dokkum et~al.,}{Van~Dokkum
  et~al.}{2008}]{van2008confirmation}
Van~Dokkum P.~G.,  et~al., 2008, The Astrophysical Journal Letters, 677, L5

\bibitem[\protect\citeauthoryear{Wellons \& Torrey}{Wellons \&
  Torrey}{2017}]{wellons2017improved}
Wellons S.,  Torrey P.,  2017, Monthly Notices of the Royal Astronomical
  Society, 467, 3887

\bibitem[\protect\citeauthoryear{Whitaker, Kriek, Van~Dokkum, Bezanson,
  Brammer, Franx  \& Labb{\'e}}{Whitaker et~al.}{2012}]{whitaker2012large}
Whitaker K.~E.,  Kriek M.,  Van~Dokkum P.~G.,  Bezanson R.,  Brammer G.,  Franx
  M.,   Labb{\'e} I.,  2012, The Astrophysical Journal, 745, 179

\bibitem[\protect\citeauthoryear{{Willett} et~al.,}{{Willett}
  et~al.}{2013}]{2013MNRAS.435.2835W}
{Willett} K.~W.,  et~al., 2013, \mn@doi [\mnras] {10.1093/mnras/stt1458}, \href
  {http://adsabs.harvard.edu/abs/2013MNRAS.435.2835W} {435, 2835}

\bibitem[\protect\citeauthoryear{Yang, Mo, Van~den Bosch, Pasquali, Li  \&
  Barden}{Yang et~al.}{2007}]{yang2007galaxy}
Yang X.,  Mo H.,  Van~den Bosch F.~C.,  Pasquali A.,  Li C.,   Barden M.,
  2007, The Astrophysical Journal, 671, 153

\bibitem[\protect\citeauthoryear{Yoon \& Park}{Yoon \&
  Park}{2020}]{yoon2020dependence}
Yoon Y.,  Park C.,  2020, The Astrophysical Journal, 897, 121

\bibitem[\protect\citeauthoryear{Yoon, Im  \& Kim}{Yoon
  et~al.}{2017}]{yoon2017massive}
Yoon Y.,  Im M.,   Kim J.-W.,  2017, The Astrophysical Journal, 834, 73

\bibitem[\protect\citeauthoryear{Zahid, Geller, Fabricant  \& Hwang}{Zahid
  et~al.}{2016}]{zahid2016scaling}
Zahid H.~J.,  Geller M.~J.,  Fabricant D.~G.,   Hwang H.~S.,  2016, The
  Astrophysical Journal, 832, 203

\bibitem[\protect\citeauthoryear{Zahid, Sohn  \& Geller}{Zahid
  et~al.}{2018}]{zahid2018stellar}
Zahid H.~J.,  Sohn J.,   Geller M.~J.,  2018, The Astrophysical Journal, 859,
  96

\bibitem[\protect\citeauthoryear{{de Carvalho}, {Ribeiro}, {Stalder}, {Rosa},
  {Costa}  \& {Moura}}{{de Carvalho} et~al.}{2017}]{decarvalho}
{de Carvalho} R.~R.,  {Ribeiro} A.~L.~B.,  {Stalder} D.~H.,  {Rosa} R.~R.,
  {Costa} A.~P.,   {Moura} T.~C.,  2017, \mn@doi [\aj]
  {10.3847/1538-3881/aa7f2b}, \href
  {https://ui.adsabs.harvard.edu/abs/2017AJ....154...96D} {154, 96}

\bibitem[\protect\citeauthoryear{de~los Rios, Mart{\'\i}nez, Coenda, Muriel,
  Ruiz, Vega-Mart{\'\i}nez  \& Cora}{de~los Rios et~al.}{2021}]{de2021roger}
de~los Rios M.,  Mart{\'\i}nez H.~J.,  Coenda V.,  Muriel H.,  Ruiz A.~N.,
  Vega-Mart{\'\i}nez C.~A.,   Cora S.~A.,  2021, Monthly Notices of the Royal
  Astronomical Society, 500, 1784

\bibitem[\protect\citeauthoryear{van~der Wel et~al.,}{van~der Wel
  et~al.}{2014}]{van20143d}
van~der Wel A.,  et~al., 2014, The Astrophysical Journal, 788, 28

\makeatother
\end{thebibliography}



\appendix

\section{Estimating the effect of contaminants in PPS regions}
\label{A}

In this appendix, we present a brief analysis on projection effects, to determine whether potential misclassifications of data points in the four regions of the PPS could lead to significant changes in the trends found for the corresponding observables. The idea is to trace such effects as clearly as possible, hence we focus on a very simple model subjected to a well-controlled bootstrap test.

\begin{figure}
\centering
\includegraphics[width=0.435\linewidth]{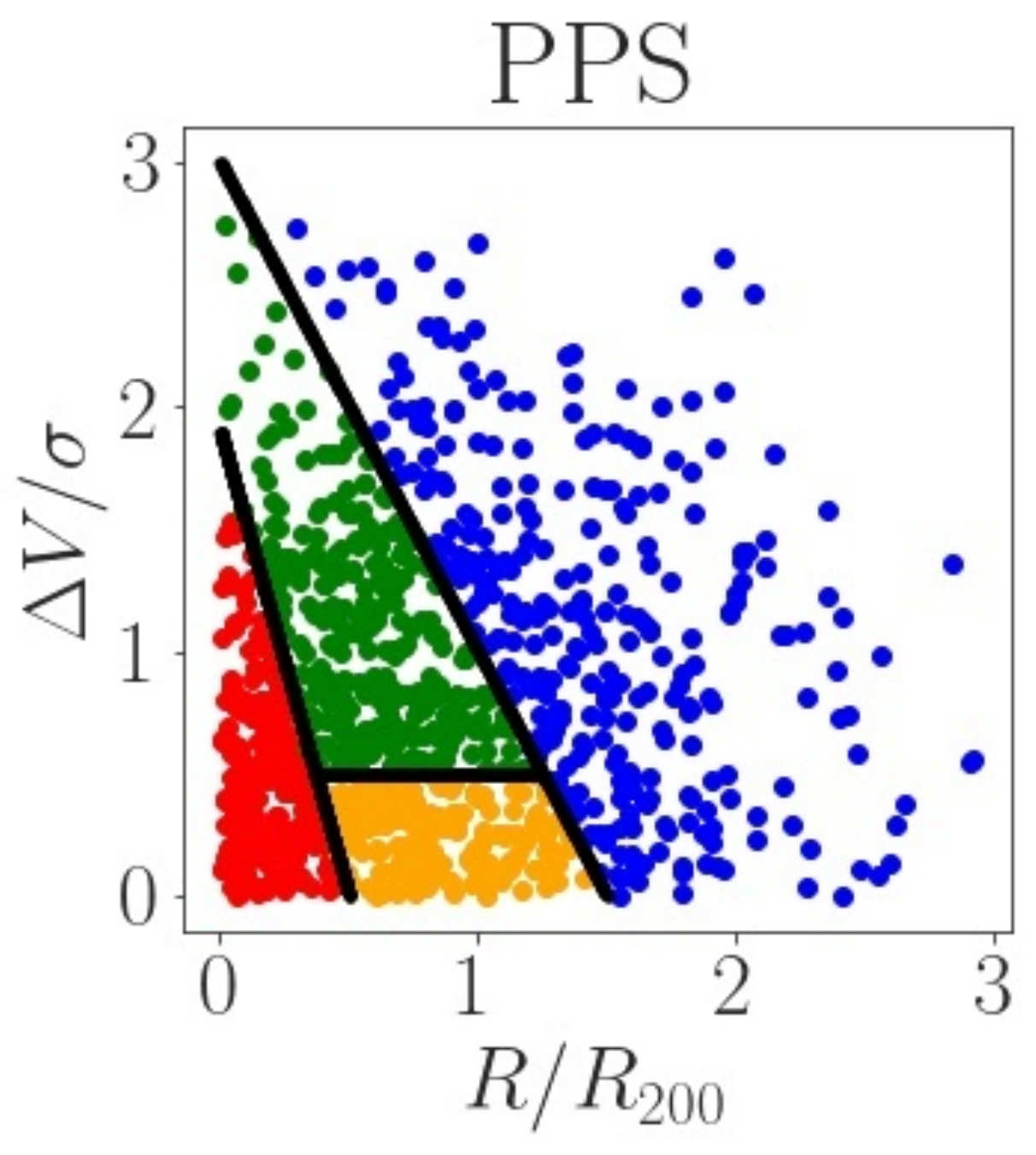}
\includegraphics[width=0.475\linewidth]{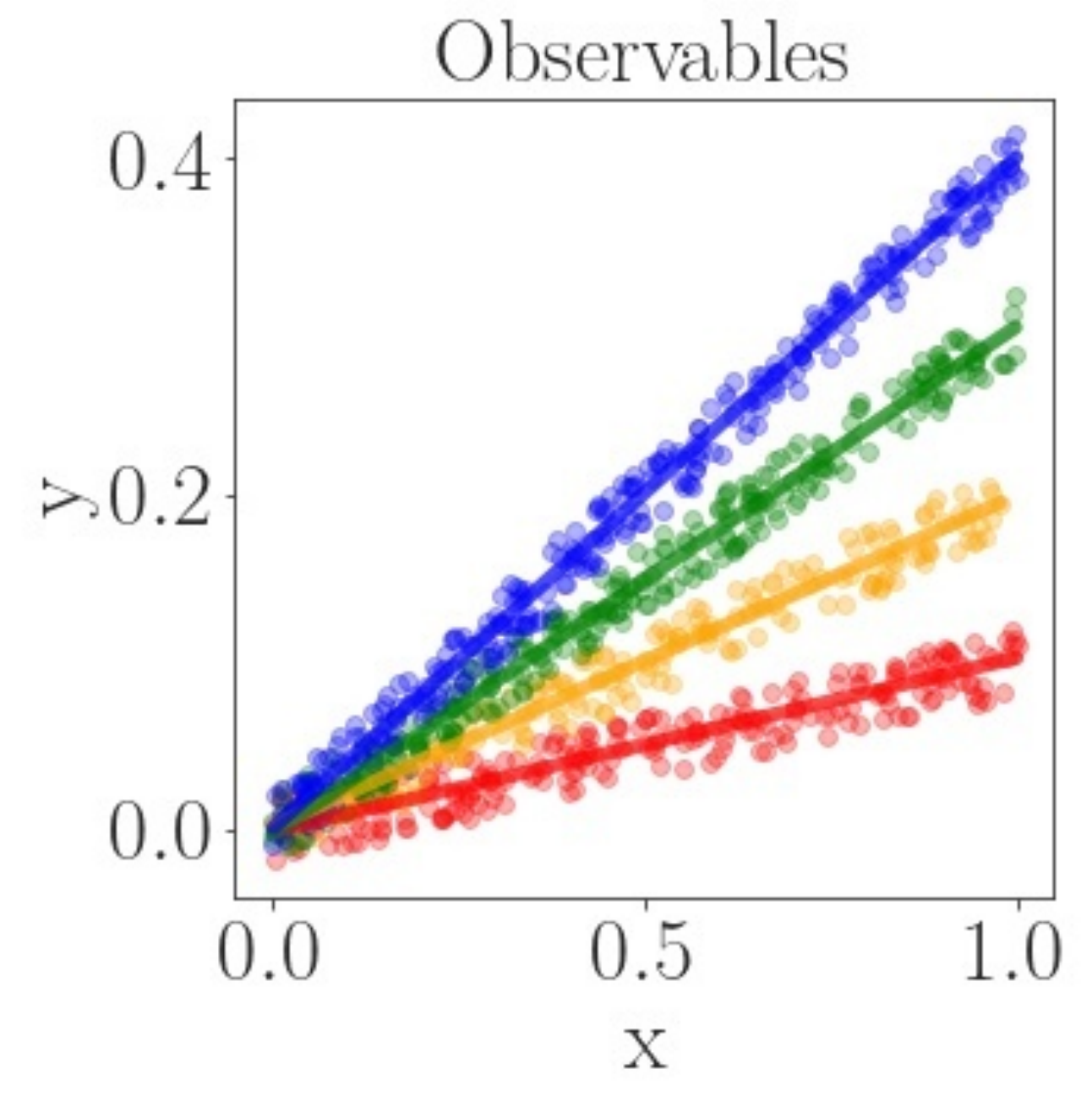}
\caption{Synthetic data set with $1000$ points. {\it Left:} PPS generated from a 2D Gaussian
distribution; {\it right:} observable trends based on a linear model. \label{ORIG}}
\end{figure}

We constructed a synthetic data set with $1000$ randomly generated points on the PPS (Fig. \ref{ORIG}, left panel), with coordinates within $0 < R/R_{200} < 3$ and $0 < \Delta V/\sigma < 3$, taken from a $2$D Gaussian distribution with mean $(0.01, 0.01)$ and unit covariance matrix. For the synthetic observable $y$ (Fig. \ref{ORIG}, right panel), we constructed linear trends, $y = ax$, with different slopes for each PPS class, randomly choosing both the $x$ coordinate (from $0$ to $1$), as well as the spread around the linear correlation ($0 < |\Delta y| < 0.02$). The slopes were $a = \{ 0.1, 0.2, 0.3, 0.4 \}$ for the classes $\mathcal {C} = \{$ ``ancient'', ``intermediate'', ``recent'', ``first infallers'' $\}$, respectively.

We assumed the same contamination fractions in each region of the PPS as given by \cite{rhee2017phase} (c.f., their Fig. 6). For each reference class, we randomly selected points from the other three regions according to the contamination fractions for this class, and reclassified the selected data points to the given reference class. 

\begin{figure*}
\centering
\includegraphics[width=\textwidth]{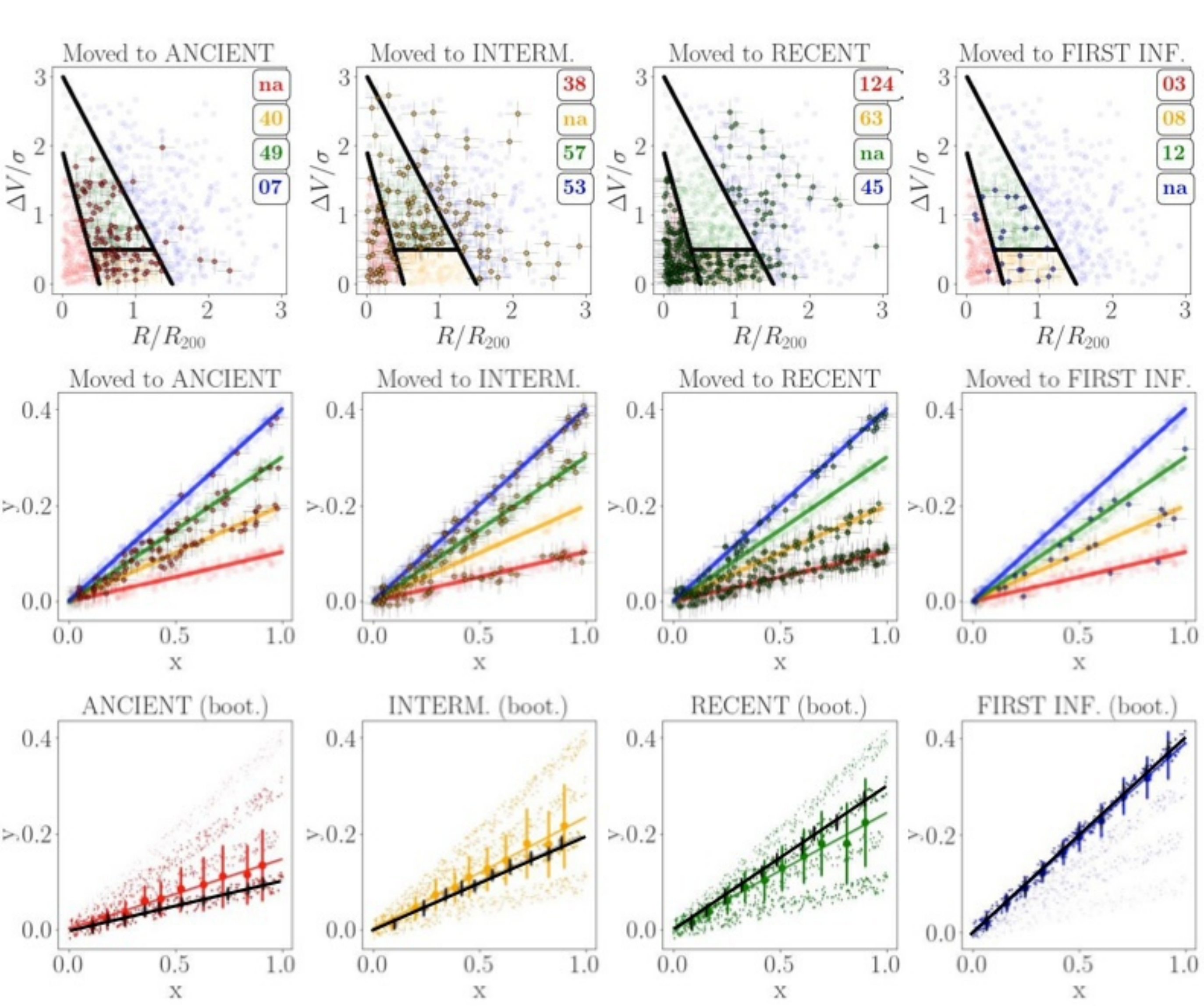}
\caption{Bootstrap analysis. {\it First row of panels:} the reclassified (or ``moved'') data points, highlighted and signaled with a cross. The number of reclassified data points for each class is shown in the legend (``na'' means not applicable).  {\it Second row of panels:} the corresponding (reclassified) data points in the observable space. {\it Third row of panels:} a linear regression fit to the binned data points generated by the bootstrap procedure. Vertical bars indicate the standard deviation of the observations in each bin. We also show the original correlations for comparison (in black).\label{PAINEIS}}
\end{figure*}

In Fig. \ref{PAINEIS} (first row of panels), we illustrate this procedure on the PPS, where the reclassified (or ``moved'') data points are highlighted and signaled with a cross. In Fig. \ref{PAINEIS} (second row of panels), we show the corresponding (reclassified) data points in the observable space.
So, for each class, we reclassified galaxies that were contaminating the other classes, 
moving them to the given class in question. 
Hence we see how the correlations of the observables change statistically.
The above procedure was independently repeated $100$ times. In Fig. \ref{PAINEIS} (third row of panels), we present for each reference class a linear regression fit to the binned data points generated by this bootstrap procedure. We also show the original correlations for comparison. We found that the impact of contamination is larger for the ranges in $x$ where the correlations of the observables differ more among classes  (namely, as $x \rightarrow 1$). The ``first infallers'' show almost no changes due to the very small contamination fractions for this class. 

It can also be observed in Fig.  \ref{PAINEIS} (first row of panels) that the overall highest numbers of misclassified data points (i.e. points that needed to be moved from other classes into a given reference class) were obtained for the ``recent'' class. This means there could be a non-negligible observational contamination of the latter class into the other classes, especially into the ``ancient'' class (i.e., $124$ data points were moved out from this class). Next in misclassification significance is the ``intermediate'' class, which required moving data points from the ``recent'' class ($57$ data points) and from the ``first infallers'' ($53$) class. Next, the ``ancient'' class received data points from the ``recent'' class ($49$) and the ``intermediate'' ($40$) class. The resulting effective contaminations can be evaluated by a comparison between the respective original correlations (black color) and those obtained after reclassifying the data points into their correct classes (c.f. the third row of panels in Fig.  \ref{PAINEIS}).

From this simple test, we obtained a qualitative gauge for the contamination effects.  It was also possible to roughly predict the direction along the $y$-axis a given class would change if misclassifications were accounted for. For instance,  the observable trend of the ``ancient'' class (Fig. \ref{PAINEIS}) moved upwards in $y$, and this is clearly because the trends of the other classes (with some misclassified ``ancients'') had higher slopes. When these misclassified ``ancients'' were corrected in the bootstrap, the tendency was to produce an ``ancient'' correlation with a higher slope. This can be similarly observed for the ``intermediate'' (upward shift) and ``recent'' classes (downward shift).

 We conclude with the following hints: (1) within some interval in $\Delta x$, a given class would probably move upwards (downwards) in $y$ depending on the corresponding values of the other classes having higher (lower) values in $y(\Delta x)$; and (2) as long as the observable correlations do not diverge too much among each other  (roughly, within their confidence intervals), the unavoidable misclassifications of classes  based on the PPS should not lead to significant changes in the trends, at least based on the contamination fractions here adopted.


\bsp	
\label{lastpage}
\end{document}